\documentclass[12pt]{article}

\usepackage[margin=1in]{geometry}
\usepackage{mathdots}
\usepackage{fancyhdr}
\usepackage{graphicx}
\usepackage{amsmath}
\usepackage{amssymb}
\usepackage{mathrsfs}
\usepackage{amsthm}
\usepackage{bbm}
\usepackage{wrapfig}
\usepackage{multicol}
\usepackage{multirow}
\usepackage{float}
\usepackage{cancel}
\usepackage{pgfplots}
\pgfplotsset{compat=1.16}
\definecolor{myblue}{rgb}{0.05,0.35,0.9}
\usepackage[colorlinks=false, hyperfootnotes=false, urlcolor=myblue, linkcolor=myblue, citecolor=myblue]{hyperref}

\usepackage[longnamesfirst]{natbib}
\usepackage[utf8]{inputenc}
\usepackage{subfiles}
\usepackage{blindtext}
\usepackage[nobottomtitles*]{titlesec}
\usepackage{lipsum}
\usepackage[title,toc,titletoc]{appendix}
\usepackage{etoolbox}
\appto\appendix{\addtocontents{toc}{\protect\setcounter{tocdepth}{1}}}
\usepackage{multicol}
\usepackage{enumerate}
\usepackage{caption}
\usepackage{subcaption}
\usepackage{graphicx}
\usepackage{tikz}
\usetikzlibrary{automata}
\usetikzlibrary{positioning}
\usetikzlibrary{arrows}
\usetikzlibrary{patterns}   
\usetikzlibrary{intersections}   
\usetikzlibrary{automata}   
\usetikzlibrary{patterns} 
\usetikzlibrary{arrows,chains,matrix,positioning,scopes}  
\usetikzlibrary{shapes,backgrounds, calc}
\usepackage[all]{xy}
\usepackage[font=footnotesize, textfont=it]{caption}

\newcommand{\RN}[1]{  \textup{\uppercase\expandafter{\romannumeral#1}}}

\newtheoremstyle{break}
  {\topsep}{\topsep}%
  {\itshape}{}%
  {\bfseries}{}%
  {\newline}{}%
\theoremstyle{break}

\newtheorem{definition}{Definition}
\newtheorem{proposition}{Proposition}
\newtheorem*{proposition*}{Proposition}

\newtheorem*{theorem*}{Theorem}
\newtheorem{corollary}{Corollary}

\theoremstyle{remark}
\newtheorem{example}{Example}
\usepackage{etoolbox}
\AtEndEnvironment{example}{\hfill$\lozenge$}
\theoremstyle{remark}

\DeclareMathOperator*{\argmax}{arg~max}
\DeclareMathOperator*{\argmin}{arg~min}

\DeclareMathOperator*{\subjecto}{subject~to}

\begin{document}
\title{Lexicographic Robustness and the Efficiency of Optimal Mechanisms\footnote{This paper subsumes ``Proper Robustness and the Efficiency of Monopoly Screening," first circulated on October 11, 2024. I thank Qingmin Liu, Marzena Rostek, and three anonymous referees for detailed feedback on the original manuscript that led to significant broadening of its scope. For encouragement and insightful remarks, I thank my colleagues at the U.S. Naval Academy, Aislinn Bohren, Tilman Börgers, Benjamin Brooks, Hector Chade, Yeon-Koo Che, Tommaso Denti, Francesc Dilmé, Laura Doval, Navin Kartik, Peter Klibanoff, Rohit Lamba, Elliot Lipnowski, Andrew Mackenzie, George Mailath, Alessandro Pavan, Doron Ravid, Juuso Toikka, Rakesh Vohra, Mark Whitmeyer, and Kun Zhang. Kai Hao Yang expertly discussed the paper at the 2026 ASSA Annual Meeting in the ``Robustness and Mechanism Design" Econometric Society session organized by Ian Ball. Refine.ink and GPT-6 Astra were used to proofread the paper. The views expressed here are solely my own and do not in any way represent the views of the U.S. Naval Academy, U.S. Navy, or the Department of Defense.}}

\author{Ashwin Kambhampati\footnote{
	Department of Economics, United States Naval Academy; \url{kambhamp@usna.edu}.}}
\date{\today}
\maketitle

\begin{abstract}
The maxmin criterion is commonly used to formalize a mechanism designer's concern for robustness over a set of possible environments. Without substantially restricting this set, however, the criterion may identify a large and economically uninformative collection of mechanisms. This paper proposes a lexicographic, belief-based approach to refining the maxmin criterion and uses it to analyze the efficiency of optimal mechanisms. In canonical screening and auction models, the strongest refinement --- proper robustness --- selects ex post efficient mechanisms. By contrast, in a public good provision model, it identifies a unique mechanism whose inefficiency may become severe in large economies.\\


\end{abstract}

\thispagestyle{empty}
\newpage
\tableofcontents

\thispagestyle{empty}
\newpage
\pagenumbering{arabic}

\section{Introduction}

A growing literature in mechanism design uses the maxmin criterion to identify mechanisms whose performance is robust over a set of possible environments. For example, a monopolist may design a menu of products to maximize its minimum profit over a set of buyer preference types; an auctioneer may design an auction to maximize her minimum revenue over a set of joint distributions over valuations; and a firm may design a production and transfer scheme to maximize its minimum profit over a set of possible externalities affecting nearby residents.

In the absence of substantial restrictions on the set of uncertainty, however, the maxmin criterion may leave important choices unresolved. In the applications discussed, it can select a large set of mechanisms, including mechanisms that are weakly dominated and that do not reflect cautiousness outside of worst-case scenarios. To address these limitations, this paper proposes a lexicographic, belief-based approach to refining the maxmin criterion and uses it to characterize the efficiency of optimal mechanisms. Under the strongest proposed refinement, ex post efficiency emerges in screening and auction environments, in contrast to the distortions emphasized in the Bayesian analysis \citep{mussa1978monopoly, myerson1981optimal}. On the other hand, substantial inefficiencies can arise under private provision of public goods, in coherence with that analysis \citep{rob1989pollution,mailath1990asymmetric}. 

The approach to modeling uncertainty is based on the non-Archimedean variant of subjective expected utility introduced by \cite{blume1991lexicographic} to characterize equilibrium refinements (see \cite{blume1991eqlexicographic}). Specifically, the designer is assumed to evaluate mechanisms using an ordered collection of beliefs called a lexicographic probability system (LPS). The first belief captures the designer’s primary assessment. Higher-order beliefs are then used to rank mechanisms that perform equally well according to lower-order beliefs.

Three increasingly strong notions of robustness are introduced. First, a mechanism is \textit{robust} if it is optimal with respect to an adversarial LPS --- an LPS whose first belief only contains worst-case type profiles in its support. Second, a mechanism is \textit{perfectly robust} if it is optimal with respect to an adversarial LPS with full support --- an adversarial LPS for which every type profile appears in the support of some belief in the LPS. Third, a mechanism is \textit{properly robust} if it is optimal with respect to a full support LPS that is also \textit{strongly} adversarial --- a full support LPS for which type profiles that are more harmful to the designer appear earlier in the belief hierarchy.

The approach has three noteworthy properties in finite settings. First, because the restrictions on the designer's LPS parallel those justifying Nash (\cite{nash1950equilibrium}), perfect (\cite{Selten_1975}) and proper (\cite{myerson1978refinements}) equilibrium strategies, robust mechanisms are optimal against a specific worst-case conjecture; perfect mechanisms are further optimal against a tremble away from a worst-case conjecture; and properly robust mechanisms are further optimal against a tremble away from a worst-case conjecture in which lower-payoff type profiles are much more likely than higher-payoff type profiles. Second, whenever a mechanism is optimal with respect to an LPS, it is optimal in the limit of a specific sequence of Bayesian priors, thereby facilitating comparison to the large literature on Bayesian mechanism design. Third, the framework nests existing refinements of maxmin optimality: robustness is equivalent to maxmin optimality; perfect robustness is equivalent to maxmin optimality and admissibility \citep{borgers2017no}; and proper robustness is equivalent to leximin optimality \citep{Rawls1971,Sen1970}.

After introducing the modeling framework and optimality criteria, the paper analyzes the efficiency implications of the refinements in several applications. The analysis first considers a screening model and a single-unit auction model. In both settings, robustness alone has limited bite, whereas perfect robustness yields sharp predictions in two-type settings. With more than two types, however, perfect robustness permits a range of allocations to intermediate types, including those exhibiting substantial distortions. By contrast, in the screening model, there is a unique properly robust mechanism. It is efficient and maximal: the allocation rule is ex post efficient and implemented by the revenue-maximal transfer rule compatible with that allocation rule. Likewise, in the auction model, proper robustness selects efficient and maximal mechanisms. It also refines the tie-breaking rule when multiple bidders have the same maximal bid: ties must be broken uniformly whenever the maximal bid is less than the highest feasible bid.

The economic force behind the efficiency results is the tension between the incentive to extract rent from higher type profiles and the offsetting uncertainty aversion captured by a strongly adversarial LPS. In both the screening and auction settings, the incentive constraints place monotonicity requirements on the allocation rules the designer implements. Revenue-maximizing transfer rules that implement such allocation rules involve binding downward-adjacent incentive compatibility constraints. Thus, distortions at lower type profiles are attractive because they reduce the information rent yielded at higher type profiles. If increasing allocations imply that the designer’s payoff is increasing in type profiles in a suitable sense, then a strongly adversarial LPS prioritizes lower type profiles over higher ones. Hence, mechanisms that reduce the designer’s payoff at lower profiles in order to extract more surplus at higher profiles are suboptimal. Efficiency at the bottom thus cascades upward, resulting in an efficient allocation rule.

To illustrate how the efficiency result breaks when increasing allocations do not imply increasing designer payoffs, a model in which a profit-maximizing firm produces a public good is considered. In the model, each agent has either a high valuation or a low valuation, and it is efficient to produce a costly public good if and only if the sum of valuations is at least the cost of production. The incentive compatibility constraints again imply that the probability of public good provision must increase in each agent's type. But if there are sufficiently many high types, each high-type agent knows that their report does not influence the decision to produce the public good under the efficient allocation rule. Hence, the designer cannot extract any more revenue from them than she could if they were a low type. Robustness considerations thus lead a profit-maximizing designer to downwardly distort the probability with which the public good is produced at lower type profiles to reduce information rents at higher type profiles, which are lexicographically salient precisely because they yield low revenue absent such distortions. Strikingly, as the number of agents grows large and the per-capita cost of providing the public good is held constant, the public good is provided with appreciable probability only when nearly all agents have high valuations. Thus, properly robust mechanisms become severely inefficient.

The rest of the paper proceeds as follows. Section \ref{sec_model} introduces the general modeling framework. Sections \ref{sec_screen}, \ref{sec_auction}, and \ref{sec_pub} analyze the screening, auction, and public good models. Section \ref{discuss} concludes and discusses related literature. A reader interested in a particular application may skip directly to the relevant section upon review of Section \ref{sec_model}. Proofs of all results are in Appendix \ref{app_proofs}. Supplemental Appendix \ref{app_examples} contains additional examples and supplementary calculations.

\section{Modeling framework}\label{sec_model}

This section describes an abstract decision problem, with mechanism design terminology used only to make clear how it maps to the applications. There is a non-empty, finite set of type profiles $\Theta$ with $N$ elements and a topological space of mechanisms $\mathcal{M}$. The principal's utility under mechanism $m \in \mathcal{M}$ and type profile $\theta \in \Theta$ is determined by the measurable function $v:\mathcal{M} \times \Theta \rightarrow \mathbb{R}$. The principal has access to Borel probability measures (lotteries) on $\mathcal{M}$ that yield finite expectations with respect to $v(\cdot, \theta)$ for all $\theta \in \Theta$. Let the set of such probability measures be denoted by $\Delta(\mathcal{M})$ and let $V: \Delta(\mathcal{M}) \times \Theta \rightarrow \mathbb{R}$ denote the extension of $v$ taking expectations over $\mathcal{M}$. Equip $\Delta(\mathcal{M})$ with a topology making the map $\sigma \mapsto V(\sigma, \theta)$ continuous for each $\theta \in \Theta$. Let $\Delta(\Theta)$ denote the space of probability measures on $\Theta$ equipped with its usual Euclidean topology.

\subsection{Lexicographic expected utility} \label{sec_lex_Bayes}

The principal possesses a lexicographic probability system and ranks mechanisms using a corresponding vector of expected utilities. A \textbf{lexicographic probability system (LPS)} is a vector of beliefs $\mu=(\mu_1, \ldots, \mu_K)$, where $K$ is a strictly positive integer and $\mu_k \in \Delta(\Theta)$ for each $k \in \{1, \ldots, K\}$. Fixing an LPS $\mu=(\mu_1, \ldots, \mu_K)$, each mechanism $\sigma \in \Delta(\mathcal{M})$ gives rise to a $K$-vector of expected payoffs: the principal's \textbf{$k$-th order payoff} from $\sigma \in \Delta(\mathcal{M})$ is
\[ \sum_{\theta \in \Theta} \mu_k(\theta) 
 V(\sigma, \theta).\]
Then, a mechanism $\sigma \in \Delta(\mathcal{M})$ is \textbf{$\mu$-optimal} if, for all $\sigma' \in \Delta(\mathcal{M})$,
\[ \left( \sum_{\theta \in \Theta} \mu_k(\theta) 
 V(\sigma, \theta) \right)^K_{k=1} \geq_{L} \left( \sum_{\theta \in \Theta} \mu_k(\theta) 
 V(\sigma', \theta) \right)^K_{k=1} ,\]
where $\geq_L$ is the lexicographic order.\begin{footnote}{For $a, b \in \mathbb{R}^K$, $a \geq_L b$ if whenever $b_k>a_k$, there exists a $j<k$ such that $a_j > b_j$.}\end{footnote}  

 Due to the relationship between LPSs and the vanishing ``trembles" used in the literature on equilibrium refinements, there is a precise sense in which $\mu$-optimal mechanisms outperform others in the limit of particular sequences of Bayesian priors converging to the first-order belief $\mu_1$. Specifically, given any LPS $\mu=(\mu_1, \ldots, \mu_K)$ and vector $r \in (0,1)^{K-1}$, define a probability measure on the set of type profiles $\Theta$ by the nested convex combination
\[ r \square \mu := (1-r_1) \mu_1 + \] \[r_1 \left[ (1-r_2) \mu_2 + r_2 \left[ (1-r_3) \mu_3 + r_3 \left[ \cdots +r_{K-2} \left[ (1-r_{K-1}) \mu_{K-1}+r_{K-1} \mu_K \right] \cdots \right] \right]  \right],\] where $r \square \mu :=\mu_1$ when $K=1$ by convention. The following result relates $\mu$ to $r \square \mu$ and will be useful to provide ``Bayesian" intuitions in applications. The proof is immediate from Proposition 4 of \cite{mailath1997proper} and therefore omitted.

\begin{proposition}[Relationship to Bayesian optimality]\label{lex_Bayes}
A mechanism $\sigma \in \Delta(\mathcal{M})$ is $\mu$-optimal if and only if for any mechanism $\sigma' \in \Delta(\mathcal{M})$ that is not $\mu$-optimal and any sequence $(r_\ell)_\ell \in ((0,1)^{K-1})^{\mathbb{N}}$ with $r_\ell \rightarrow (0, 0, \ldots, 0)$, there exists an $L \in \mathbb{N}$ such that
\[ \sum_{\theta \in \Theta} (r_\ell \square \mu)(\theta) V(\sigma, \theta) > \sum_{\theta \in \Theta} (r_\ell \square \mu)(\theta) 
 V(\sigma', \theta) \quad \text{for all $\ell \geq L$.}\]
\end{proposition}

\subsection{Robust mechanisms}\label{framework_robust}

It will be of interest to restrict the principal's LPS and study the implications of these restrictions on the form of lexicographically optimal mechanisms. The first such restriction requires that the support of the first belief only contains worst-case type profiles.

\begin{definition}
An LPS $\mu=(\mu_1, \ldots, \mu_K)$ is \textbf{adversarial} with respect to $\sigma \in \Delta(\mathcal{M})$ if $\mu_1(\theta)>0$ implies
\[V(\sigma, \theta)  \leq V(\sigma, \theta') \quad \text{for all $\theta' \in \Theta$.}\]
\end{definition}

A robust mechanism is one that can be rationalized by some adversarial LPS.

\begin{definition}
   A mechanism $\sigma \in \Delta(\mathcal{M})$ is \textbf{robust} if there exists an LPS, $\mu$, that is adversarial with respect to $\sigma$ and for which $\sigma$ is $\mu$-optimal. 
\end{definition}

Exploiting the minimax theorem, the following proposition confirms that robustness is equivalent to maxmin optimality.

 \begin{proposition}[Robustness and maxmin criterion]\label{robust_maxmin}
A mechanism $\sigma \in \Delta(\mathcal{M})$ is robust if and only if it is maxmin optimal, i.e., the mechanism $\sigma \in \Delta(\mathcal{M})$ solves the optimization problem
\begin{equation}
\max_{\sigma' \in \Delta(\mathcal{M})}  \min_{\theta' \in \Theta}  V(\sigma', \theta'). \notag
\end{equation} 
\end{proposition}

\subsection{Perfectly robust mechanisms}\label{model_perfect}

Maxmin optimal mechanisms may not be admissible.

\begin{definition}
	A mechanism $\sigma' \in \Delta(\mathcal{M})$ \textbf{weakly dominates} the mechanism $\sigma \in \Delta(\mathcal{M})$ if \[V(\sigma', \theta) \geq V(\sigma, \theta) \quad \text{for all $\theta \in \Theta$}\] and the inequality is strict for some $\theta \in \Theta$. A mechanism $\sigma \in \Delta(\mathcal{M})$ is \textbf{admissible} if there does not exist a mechanism $\sigma' \in \Delta(\mathcal{M})$ that weakly dominates it.
\end{definition}

The next LPS restriction will be sufficient to rule out inadmissible mechanisms.

\begin{definition}
An LPS $\mu=(\mu_1, \ldots, \mu_K)$ has \textbf{full support} if, for each $\theta \in \Theta$, there exists a $k \in \{1, \ldots, K\}$ such that $\mu_k(\theta)>0$.
\end{definition}

\noindent A mechanism is defined to be perfectly robust if it is optimal against a full support LPS that is adversarial.

\begin{definition}
A mechanism $\sigma \in \Delta(\mathcal{M})$ is \textbf{perfectly robust} if there exists a full support LPS, $\mu$, that is adversarial with respect to $\sigma$ and for which $\sigma$ is $\mu$-optimal.
\end{definition}

The following relationship between perfect robustness and the admissibility refinement of the maxmin criterion holds.

\begin{proposition}[Perfect robustness and admissibility]\label{perfect_admissibility}
If a mechanism is perfectly robust, then it is maxmin optimal and admissible. The converse holds if $\mathcal{M}$ is finite: If $|\mathcal{M}|<\infty$ and a mechanism is both maxmin optimal and admissible, then it is perfectly robust.
\end{proposition}

The proof of the converse uses finiteness to ensure that every
admissible mechanism is a best response to some full-support prior.
This property can fail for infinite mechanism sets even when
$\mathcal M$ is compact and $v(\cdot,\theta)$ is continuous.
See, for instance, Example 2 in Section 2.2 of
\cite{ChengBorgers_2025}.

\subsection{Properly robust mechanisms}\label{model_prop}

The final restriction on beliefs ensures that the principal's higher-order uncertainty is consistent with her lower-order uncertainty. To define these restrictions, it will be useful to introduce an order on type profiles. Given an LPS $\mu$ and type profiles $\theta, \theta' \in \Theta$, write $\theta >_\mu \theta'$ and say that $\theta$ is ``infinitely more likely" than $\theta'$ if
\[ \min\{k \in \{1, \ldots, K\}: \mu_k(\theta)>0\}< \min\{k \in \{1, \ldots, K\}:\mu_k(\theta')>0\}, \] where, by convention, $\min \emptyset=+\infty$. The weak order is defined in the usual way: if it is not the case that $\theta' >_\mu \theta$, then $\theta \geq_\mu \theta'$. A strongly adversarial LPS is then defined as one in which lower-payoff type profiles are infinitely more likely than higher-payoff type profiles.

\begin{definition}
An LPS $\mu=(\mu_1, \ldots, \mu_K)$ is \textbf{strongly adversarial} with respect to $\sigma \in \Delta(\mathcal{M})$ if, for all $\theta \in \Theta$,
\[V(\sigma, \theta)  \leq V(\sigma, \theta')  \quad  \text{for all $\theta' \in \Theta$ with $\theta \geq_\mu \theta'$.}\]
\end{definition}

Notice that, if $\mu_1(\theta)>0$, then $\theta \geq_\mu \theta'$ for all $\theta' \in \Theta$. So, if $\mu$ is strongly adversarial, then it is adversarial:
\[V(\sigma, \theta)  \leq V(\sigma, \theta') \quad \text{for all $\theta' \in \Theta$.}\] A mechanism is defined to be properly robust if it is optimal against a full support LPS that satisfies the stronger requirement.

\begin{definition}
A mechanism $\sigma \in \Delta(\mathcal{M})$ is \textbf{properly robust} if there exists a full support LPS, $\mu$, that is strongly adversarial with respect to $\sigma$ and for which $\sigma$ is $\mu$-optimal.
\end{definition}

It is next shown that proper robustness is a belief-based analog of the leximin criterion. To define the leximin criterion in the current setting, note that the mechanism $\sigma \in \Delta(\mathcal{M})$ is associated with an $N$-vector of expected payoffs $(V(\sigma,\theta))_{\theta \in \Theta}$. Let $V_{(i)}(\sigma)$ denote the $i$-th order statistic of this vector, i.e., the $i$-th lowest value in $(V(\sigma,\theta))_{\theta \in \Theta}$. Then, a mechanism $\sigma \in \Delta(\mathcal{M})$ is \textbf{leximin optimal} if, for all $\sigma' \in \Delta(\mathcal{M})$,
\[ \left( V_{(i)}(\sigma) \right)^N_{i=1} \geq_{L} \left( V_{(i)}(\sigma') \right)^N_{i=1} ,\]
where $\geq_L$ is the lexicographic order. The characterization follows below.

\begin{proposition}[Proper robustness and leximin optimality]\label{proper_leximin}
If a mechanism is properly robust, then it is leximin optimal. The converse holds if $\mathcal{M}$ is finite: If $|\mathcal{M}|<\infty$ and a mechanism is leximin optimal, then it is properly robust.
\end{proposition}

Proposition \ref{proper_leximin} is of independent theoretical interest because it provides an alternative characterization of the leximin criterion. Modifying \cite{savage1954foundations}'s framework to allow for discontinuous preferences over subjective lotteries, \cite{Mackenzie_2024} recently observed that the leximin criterion reflects an axiom of maximal risk aversion in an appropriately defined subjective lexicographic expected utility framework.\footnote{The leximin criterion may also be interpreted as a lexicographically ordered form of downside-risk aversion \citep{manski1988ordinal, rostek2010quantile}.} \cite{Mackenzie_2024} remarks that there is a sense in which the subjective lexicographic expected utility model he studies (which involves a unique probability measure and lexicographic utilities) is the ``dual" of the lexicographic expected utility model of \cite{blume1991lexicographic} (which involves a unique utility function and lexicographic probabilities). Proper robustness links the two approaches.

\subsection{Remarks about the criteria}

It is immediate from the definitions that every properly robust mechanism is perfectly robust and every perfectly robust mechanism is robust. Note that, if a mechanism is robust, then it is a part of a saddle point in a fictitious zero-sum game between the principal and an adversarial opponent (``Nature") that chooses the measure over type profiles. The restrictions on the principal's LPS sustaining a perfectly robust mechanism correspond to those that sustain a (trembling-hand) perfect equilibrium strategy in this game. Precisely, if a decision-maker's strategy is a part of a perfect equilibrium in a finite game, then she must play a strategy that is optimal with respect to a full support LPS whose first-order belief coheres with the other's chosen strategy (see Proposition 4 of \cite{blume1991eqlexicographic}). Similarly, the restrictions on the principal's LPS sustaining a properly robust mechanism correspond to those that sustain a proper equilibrium strategy. In particular, if a decision-maker's strategy is a part of a proper equilibrium in such a game, then she must play a strategy that is optimal with respect to a full support LPS whose higher-order beliefs respect the preferences of the other player (see Proposition 5 of \cite{blume1991eqlexicographic}). In a two-player, zero-sum game, the restriction to preference-respecting LPSs corresponds to the requirement that the principal play a best response to a strongly adversarial LPS.\footnote{\cite{melvin1961games} (Chapter 3, Section 18) proposes an approach to selecting strategies in two-player, zero-sum games that involves maximizing the (guaranteed) gain resulting from a strategic opponent's ``mistakes". \cite{van1983refinements} observes that this procedure identifies proper equilibrium strategies in such games (see Chapter 3, Theorem 3.5.5). Choosing an optimal mechanism with respect to the criterion considered in this paper thus coheres with what would arise following the protocol of \cite{melvin1961games}.}

It is immediate from the preceding observations that, if the set of mechanisms is assumed to be finite (or discretized), then the sets of robust, perfect, and proper mechanisms are non-empty by the existence of Nash, perfect, and proper equilibria \citep{nash1950equilibrium,Selten_1975,myerson1978refinements}. Note, however, that perfection or properness of a mechanism does not immediately guarantee that it is played in some perfect or proper equilibrium.  The reason is that no restrictions are placed on the perfection or properness of the measure chosen by Nature in response to the principal's mechanism. This is a natural relaxation given that Nature is an artificial player used to facilitate understanding of the principal's attitude towards uncertainty. 

In the applications that follow, the necessary directions in Propositions \ref{perfect_admissibility} and \ref{proper_leximin} are used to derive necessary conditions for optimality. The converse directions, by contrast, do not apply as stated to the applications in Sections \ref{sec_screen}, \ref{sec_auction}, and \ref{sec_pub} because the sets of mechanisms are infinite. Instead, optimality is established by constructing justifying LPSs, which in turn yield Bayesian foundations through Proposition \ref{lex_Bayes}.

\section{Screening}\label{sec_screen}

This section analyzes a general, single-agent screening model. While the model is described using the product design language of \cite{mussa1978monopoly}, it also nests (discrete-type versions of) the single-agent pricing model of \cite{riley1983optimal}, the nonlinear pricing model of \cite{maskin1984monopoly}, and, with some transformations, the monopoly regulation model of \cite{baron1982regulating}. The model is simplified and extended to multi-agent settings in Sections \ref{sec_auction} and \ref{sec_pub}.

\subsection{Environment}\label{sec_scree_enviro}
The principal is a seller who transacts with a buyer. The seller chooses a quality $q \in \mathcal{Q}$, where $\mathcal{Q} \subseteq \mathbb{R}_+$ contains zero, at a cost determined by the function $c: \mathcal{Q} \rightarrow \mathbb{R}_+$. The seller's ex post payoff from selling a good of quality $q \in \mathcal{Q}$ at a price of $p \in \mathbb{R}$ is given by
\[p-c(q). \] The buyer has private information about her payoff type $\theta \in \Theta$, where $\Theta := \{\theta_1, \ldots, \theta_N\} \subseteq \mathbb{R}$ and $ 0<\theta_1 < \cdots < \theta_N$. Specifically, the buyer's ex post payoff from purchasing a good of quality $q \in \mathcal{Q}$ at a price of $p \in \mathbb{R}$ is
\[ u(q,\theta)- p,\] where $u: \mathcal{Q} \times \Theta \rightarrow \mathbb{R}_+$ satisfies $u(0,\theta)=0$ for all $\theta \in \Theta$. Moreover, $u$ is increasing in $q$ and has strictly increasing differences: if $q'>q$ and $\theta'>\theta$, then
\[ u(q', \theta')- u(q, \theta') > u(q',\theta)-u(q,\theta).\] The buyer's payoff from not transacting with the seller is the same for all types and normalized to zero. For simplicity, it is assumed that there exists a unique efficient quality for each type: for each $i=1, \ldots, N$,
\[ \{ q^*_i \} :=\argmax_{q \in \mathcal{Q}} \quad u(q,\theta_i)- c(q).\] This property holds if, for instance, $\mathcal{Q}$ is a compact interval, $u$ and $c$ are continuous, and the surplus function $u( \cdot, \theta_i)-c(\cdot)$ is strictly quasiconcave for each $i=1, \ldots, N$.

The seller has commitment power and can sell different qualities at different prices. By the Revelation Principle, without loss of optimality, the seller can restrict attention to direct mechanisms.  A \textbf{(direct) mechanism} is an allocation rule and transfer rule,
\begin{equation}
Q: \Theta \rightarrow \Delta(\mathcal{Q}) \quad \text{and} \quad P: \Theta \rightarrow \mathbb{R},\notag
\end{equation}
that together are \textbf{incentive compatible} and \textbf{individually rational}: for all $\theta \in \Theta$, 
\[ U(Q(\theta), \theta)-P(\theta) \geq  U(Q(\theta'), \theta)-P(\theta') \quad \text{for all $\theta' \in \Theta$} \]
and
\[ U(Q(\theta), \theta)-P(\theta) \geq 0,\] where $U: \Delta(\mathcal{Q}) \times \Theta \rightarrow \mathbb{R}$ is the extension of $u$ taking expectations over allocations and $\Delta(\mathcal Q)$ denotes the Borel probability measures on $\mathcal Q$ under which $c$ and $u(\cdot,\theta)$ have finite expectations for every $\theta\in\Theta$. Let $\mathcal{M}$ denote the space of all mechanisms. Given a mechanism $(Q,P) \in \mathcal{M}$ and a type $\theta \in \Theta$, the seller obtains a payoff of
\[v((Q,P), \theta)=P(\theta)-C(Q(\theta)), \]
where $C: \Delta(\mathcal{Q}) \rightarrow \mathbb{R}$ is the extension of $c$ taking expectations over allocations. It can easily be seen that any randomization over mechanisms $\sigma \in \Delta(\mathcal{M})$ is payoff equivalent to some mechanism $(Q,P) \in \mathcal{M}$. Henceforth, with no loss of generality, the seller uses a mechanism $(Q,P) \in \mathcal{M}$ with the understanding that each $(Q,P)$ is technically a class of payoff equivalent mechanisms.

Any mechanism $(Q,P) \in \mathcal{M}$ in which, for all $\theta \in \Theta$, $Q(\theta)$ is the Dirac measure on some quality $q \in \mathcal{Q}$ is called a \textbf{deterministic mechanism}. An allocation $Q(\theta_i) \in \Delta(\mathcal{Q})$ is \textbf{efficient for type $\theta_i$} if $Q(\theta_i)$ is the Dirac measure on the efficient quality $q^*_i$. An allocation rule $Q: \Theta \rightarrow \Delta(\mathcal{Q})$ is \textbf{efficient} if it is efficient for all types $\theta \in \Theta$. Because $u$ satisfies strictly increasing differences, the Monotone Selection Theorem (\cite{milgrom1994monotone}) ensures that efficient allocations are increasing in type: $q^*_1 \leq q^*_2 \leq \cdots \leq q^*_N$. Hence, the unique efficient allocation rule is implementable (\cite{rochet1987necessary}). Given an efficient allocation rule, the pointwise revenue maximizing prices are characterized by the binding downward-adjacent incentive compatibility constraints and the individual rationality constraint for the lowest type. Specifically, the mechanism $(Q,P) \in \mathcal{M}$ is \textbf{efficient and maximal} if and only if $Q$ is efficient and $P$ satisfies
\begin{equation}
\begin{aligned}
P(\theta_1) &= U(Q(\theta_1), \theta_1) \quad \text{and} \\
P(\theta_j) &= U(Q(\theta_1), \theta_1)+\sum^j_{i=2} \left(U(Q(\theta_{i}), \theta_i)- U(Q(\theta_{i-1}), \theta_{i}) \right) \quad \text{for $j \in \{2,...,N\}$.}
\end{aligned}\label{rev}
\end{equation}
Finally, it will be useful to let $\delta_i \in \Delta(\Theta)$ denote the measure that places probability one on buyer type $\theta_i$.

\subsection{Robust mechanisms}

The first result establishes that few restrictions are placed on the seller's mechanism if the only requirement is that she choose a best response to an adversarial LPS; any mechanism that is ``efficient at the bottom" and does not yield excessive rent is robustly optimal.

\begin{proposition}[Screening: characterization of robust mechanisms]\label{thm_robust}
A mechanism $(Q,P) \in \mathcal{M}$ is robust if and only if
\begin{enumerate}
    \item it is efficient for the lowest type, $\theta_1$;
    \item the seller extracts full surplus from the lowest type,
\[ P(\theta_1)=u(q^*_1, \theta_1);\]
\item and the seller attains the lowest payoff from the lowest type,  \[v((Q,P),\theta) \geq v((Q,P),\theta_1) \quad \text{for all $\theta \in \Theta$}.\]
\end{enumerate}
\end{proposition}

A simple example illustrates the multiplicity of robustly optimal mechanisms.

\begin{example} \label{ex_robust}
    Suppose $\Theta=\{\theta_1, \theta_2\}$, $\mathcal{Q}= \mathbb{R}_+$, $u(q,\theta)= \theta q$, and $c(q)= \frac{1}{2} q^2$. Then, the efficient quality for type $\theta_i$ is $q^*_i := \theta_i$. Proposition \ref{thm_robust} establishes that a mechanism is robustly optimal if and only if $Q(\theta_1)$ is the Dirac measure on $\theta_1$, $P(\theta_1)=u(q^*_1,\theta_1)=\theta^2_1$,
the incentive compatibility constraints are satisfied,
\[0 \geq \theta_1 \mathbb{E}_{Q(\theta_2)}[q]- P(\theta_2) \quad \text{and} \quad \theta_2 \mathbb{E}_{Q(\theta_2)}[q]- P(\theta_2)  \geq  \theta_2 \theta_1 -\theta^2_1,  \]
and a profitability constraint is satisfied,
\[ P(\theta_2)- \frac{1}{2} \mathbb{E}_{Q(\theta_2)}[q^2]= v((Q,P), \theta_2) \geq v((Q,P), \theta_1)= \frac{1}{2} \theta^2_1.\]
The quality for type $\theta_2$ need not be efficient and can be distorted upwards or downwards. For instance, if $\theta_1=1$ and $\theta_2=2$, then any Dirac measure on $q_2 \in [1 , 3]$ is a part of a robustly optimal mechanism. Supplemental Appendix \ref{app_ex_screen_dom} shows that any robust mechanism that is \textit{not} equal to the efficient and maximal mechanism is weakly dominated.\end{example}

\subsection{Perfectly robust mechanisms}

The following result establishes that, in any setting with two types, the unique perfectly robust mechanism is the efficient and maximal mechanism. More generally, efficiency must always arise ``at the top" and ``at the bottom" in any deterministic and perfectly robust mechanism.

\begin{proposition}[Screening: characterization of perfectly robust mechanisms]\label{thm_perfect}
If there are two types ($N=2$), then the unique perfectly robust mechanism is the efficient and maximal mechanism. More generally, in any deterministic and perfectly robust mechanism $(Q,P) \in \mathcal{M}$, $Q$ is efficient for $\theta_1$ and $\theta_N$, and $P$ satisfies \eqref{rev}. 
\end{proposition}

With more than two types, perfectly robust mechanisms need not be efficient ``in the middle". The following example --- a minimal departure from Example \ref{ex_robust} --- demonstrates that intermediate types can be maximally distorted downwards subject to monotonicity constraints on the allocation rule.

\begin{example}\label{ex_properrobust}
Suppose $\Theta=\{\theta_1, \theta_2, \theta_3\}$, $\mathcal{Q}= \mathbb{R}_+$, $u(q,\theta)= \theta q$, and $c(q)= \frac{1}{2} q^2$. Consider a mechanism $(Q,P) \in \mathcal{M}$ in which both $Q(\theta_1)$ and $Q(\theta_2)$ are the Dirac measure on $q^*_1=\theta_1$ and $Q(\theta_3)$ is the Dirac measure on $q^*_3=\theta_3$. Moreover, suppose that $P$ is determined by \eqref{rev}. Supplemental Appendix \ref{app_ex_properrobust} verifies that $(Q,P) \in \mathcal{M}$ is $\mu$-optimal with respect to the adversarial and full support LPS $\mu=(\delta_1, \delta_3, \delta_2)$. Hence, $(Q,P)$ is perfectly robust. Notice that the allocation to the middle type is completely distorted downwards subject to monotonicity of the allocation rule. This distortion is optimal because profit from type $\theta_3$ is lexicographically prioritized over profit from type $\theta_2$. Hence, distorting type $\theta_2$'s allocation reduces the information rent left to type $\theta_3$.\end{example}

\subsection{Properly robust mechanisms}\label{sec_screen_prop}

Are the beliefs sustaining the inefficient mechanism in Example \ref{ex_properrobust} reasonable? If the seller were truly concerned with the robustness of her mechanism, then she might be relatively more concerned about a buyer type that is more harmful for her bottom line than a type that is less harmful. In particular, the LPS $\mu'=(\delta_1, \delta_2, \delta_3)$ seems more consistent with uncertainty aversion than $\mu=(\delta_1, \delta_3, \delta_2)$ because the seller obtains a strictly higher payoff from $\theta_3$ than $\theta_2$. The main result of the analysis of the screening model is that, when the seller must best-respond to a full support and strongly adversarial LPS (such as $\mu'=(\delta_1, \delta_2, \delta_3)$ in Example \ref{ex_properrobust}), a sharp prediction arises: the seller's unique optimal mechanism is the efficient and maximal mechanism.

\begin{proposition}[Characterization of properly robust mechanisms]\label{thm_proprobust}
The unique properly robust mechanism is the efficient and maximal mechanism.
\end{proposition}
 
Proposition \ref{thm_proprobust} implies that, in Example \ref{ex_properrobust}, the efficient and maximal mechanism is properly robust; Supplemental Appendix \ref{app_ex_properrobust_2} verifies that it is $\mu$-optimal with respect to the full support and strongly adversarial LPS $\mu=(\delta_1, \delta_2, \delta_3)$. The logic is simple: against $\delta_1$, any $\mu$-optimal mechanism must have $Q(\theta_1)$ equal to the Dirac measure on $q^*_1$ with prices extracting full surplus. Within the set of mechanisms with this property, any optimal mechanism against $\delta_2$ must set $Q(\theta_2)$ equal to the Dirac measure on $q^*_2$ with prices extracting as much surplus as possible subject to the downward-adjacent incentive compatibility constraint. Repeating the argument for the highest type establishes the result. Proposition \ref{thm_proprobust} establishes further that no
other mechanism is properly robust.

\subsection{Bayesian foundations} \label{sequence}

The proof of Proposition \ref{thm_proprobust} establishes that the unique properly robust mechanism is a best response to the LPS $\mu=(\delta_1, \delta_2, \ldots, \delta_N)$. By Proposition \ref{lex_Bayes}, it therefore outperforms other mechanisms in the limit of particular sequences of Bayesian priors converging to $\delta_1$. For simplicity of exposition, suppose that the seller restricts herself to using deterministic mechanisms and $\rho$ is a prior for which $\rho(\theta_i)>0$ for all $i$. Then, a (deterministic) mechanism is $\rho$-optimal if and only if $P$ satisfies \eqref{rev} and $Q$ solves
\begin{equation}
\begin{aligned}
&\max_{Q: \Theta \rightarrow \mathcal{Q}} \quad \sum^N_{i=1} \rho(\theta_i) \left[ u(Q(\theta_i),\theta_i)-c(Q(\theta_i))-d(\theta_i) \right] \\
&\text{subject to}\\
& d(\theta_i)=  h(\theta_i) \left( u(Q(\theta_i), \theta_{i+1})-u(Q(\theta_i),\theta_i) \right) \quad \text{for $i<N$} \\
& d(\theta_N)=0\\
& Q(\cdot)~\text{increasing,}
\end{aligned} \notag \label{Bayesian}
\end{equation}
where $h(\theta_i)=\sum^N_{j=i+1} \rho(\theta_j)/ \rho(\theta_i)$ is the inverse hazard rate for type $\theta_i$ and $d(\theta_i)$ is a term that distorts the allocation to type $\theta_i$. Suppose further that $|\Theta|=3$ and consider a sequence $(r^1_\ell, r^2_\ell)_\ell \in ((0,1)^{2})^{\mathbb{N}}$ for which $(r^1_\ell, r^2_\ell) \rightarrow (0,0)$. Then, the Bayesian prior $\rho_\ell:=(r_\ell \square \mu)$ sets
\[
\rho_\ell(\theta_1)=1-r^1_\ell,\quad
\rho_\ell(\theta_2)=r^1_\ell(1-r^2_\ell),\quad \text{and}\quad
\rho_\ell(\theta_3)=r^1_\ell r^2_\ell.
\]
The implied inverse hazard rates are
\[h_\ell(\theta_1) =\frac{r^1_\ell}{1-r^1_\ell}, \quad 
h_\ell(\theta_2)=\frac{r^2_\ell}{1-r^2_\ell},  \quad \text{and} \quad
h_\ell(\theta_3)=0.\]
If $(Q',P') \in \mathcal{M}$ is a fixed deterministic mechanism that is \textit{not} efficient and maximal, then Proposition \ref{lex_Bayes} implies that the efficient and maximal mechanism yields strictly higher expected profits than $(Q',P')$ under $\rho_\ell$ for all sufficiently large $\ell$. Hence, if lower value buyers are much more likely than higher value buyers, inverse hazard rates converge to zero. Correspondingly, the distortion terms vanish.

\section{Auction design}\label{sec_auction}

The screening model of Section \ref{sec_screen} nests a single-agent, single-unit auction model with the set of types $\Theta$ interpreted as the set of bidder valuations; set $\mathcal{Q} = \{0,1\}$, $u(q,\theta)=\theta q$, and suppose that the cost of production is constant in $q \in \mathcal{Q}$. Proposition \ref{thm_robust} then implies that the unique robust mechanism is a take-it-or-leave-it offer at the lowest type value. This section extends this model to allow for multiple bidders, as in \cite{myerson1981optimal}. This introduces significant complexity into the analysis given the large set of feasible type profiles. Nevertheless, sharp predictions arise under perfection when there are binary types, and under properness in general.

\subsection{Environment}

The principal is now an auctioneer endowed with a single indivisible good and there are $I \geq 2$ agents (bidders) $i \in \{1,2, \ldots, I\} =: \mathcal{I}$. Each agent $i$ has private information about their payoff from obtaining the good $\theta^i \in \Theta^i:= \{\theta_1, \ldots, \theta_N\} \subseteq \mathbb{R}$, where $ 0<\theta_1 < \cdots < \theta_N$ and $N \geq 2$. Let the set of type profiles be denoted by $\Theta:= \prod^I_{i=1} \Theta^i$, the minimal element of $\Theta$ be denoted by $\underline{\theta}:=(\theta_1, \ldots, \theta_1)$, and the maximal element of $\Theta$ be denoted by $\overline{\theta}:=(\theta_N, \ldots, \theta_N)$. The ex post payoff of agent $i$ with type $\theta^i$ is given by
\[q^i \theta^i-p^i, \] where $q^i \in [0,1]$ is the probability with which the agent receives the good and $p^i \in \mathbb{R}$ is the transfer from agent $i$ to the auctioneer. The auctioneer's ex post payoff is her revenue
\[\sum^I_{i=1} p^i. \]

The auctioneer has commitment power. By the Revelation Principle, without loss of optimality, she can restrict attention to direct mechanisms.  A \textbf{(direct) mechanism} consists of an allocation rule and a transfer rule. An allocation rule is a function
\begin{equation}
Q: \Theta \rightarrow [0,1]^I \quad \text{such that} \quad \sum^I_{i=1} Q^i(\theta) \leq 1,\notag
\end{equation}
where $Q^i(\theta)$ is the $i$-th component of the vector $Q(\theta)$ and denotes the probability with which agent $i$ receives the good under type profile $\theta \in \Theta$. A transfer rule is a function
\begin{equation}
P: \Theta \rightarrow \mathbb{R}^I, \notag
\end{equation}
where $P^i(\theta)$ is the $i$-th component of the vector $P(\theta)$ and denotes the transfer from agent $i$ to the auctioneer. A mechanism is \textbf{(dominant strategy) incentive compatible} and \textbf{(ex post) individually rational} if, for all $ \theta \in \Theta$ and all $i=1,2, \ldots, I$, 
\[ Q^i(\theta^i, \theta^{-i}) \theta^i-P^i(\theta^i, \theta^{-i}) \geq  Q^i(\hat{\theta}^i, \theta^{-i}) \theta^i-P^i(\hat{\theta}^i, \theta^{-i}) \quad \text{for all $\hat{\theta}^i \neq \theta^i$} \]
and
\[ Q^i(\theta^i, \theta^{-i}) \theta^i-P^i(\theta^i, \theta^{-i}) \geq 0.\] Let $\mathcal{M}$ denote the set of all mechanisms that are incentive compatible and individually rational.\footnote{As is well-known, restricting attention to dominant strategy incentive compatible and ex post individually rational mechanisms is without loss of optimality in the Bayesian independent private values model of \cite{myerson1981optimal}.} Given a mechanism $(Q,P) \in \mathcal{M}$ and a type profile $\theta \in \Theta$, the auctioneer obtains a payoff of
\[v((Q,P), \theta)= \sum^I_{i=1} P^i(\theta).\] As in the screening environment, it is easy to show that any randomization over mechanisms is payoff equivalent to a mechanism in $\mathcal{M}$. Henceforth, the auctioneer chooses an element of $\mathcal{M}$. 

The allocation rule $Q: \Theta \rightarrow [0,1]^I$ is \textbf{efficient for type profile $\theta$} if $\sum^I_{i=1} Q^i(\theta)=1$ and $Q^i(\theta)>0$ only if $\theta^i \geq \theta^j$ for all $j \neq i$. The allocation rule $Q: \Theta \rightarrow [0,1]^I$ is \textbf{efficient} if it is efficient for all type profiles $\theta \in \Theta$. The mechanism $(Q,P) \in \mathcal{M}$ is \textbf{efficient and maximal} if $Q$ is efficient and $P$ satisfies, for all $i=1,2, \ldots, I$ and all $\theta^{-i} \in \Theta^{-i}$,
\begin{equation}
\begin{aligned}
P^i(\theta_1, \theta^{-i}) &= Q^i(\theta_1, \theta^{-i}) \theta_1 \quad \text{and} \\
P^i(\theta_k, \theta^{-i}) &= Q^i(\theta_1, \theta^{-i}) \theta_1+\sum^k_{j=2} (Q^i(\theta_j,\theta^{-i})-Q^i(\theta_{j-1},\theta^{-i}) ) \theta_j \quad \text{for $k \in \{2,...,N\}$.}
\end{aligned}\label{rev_auction}
\end{equation}
In contrast to the screening model, there are multiple efficient and maximal mechanisms. Each is distinguished by the way in which ties are broken. For instance, the \textbf{efficient and maximal mechanism with uniform tie-breaking} sets
\[Q^i(\theta)= \begin{cases} \frac{1}{| \underset{j \in \mathcal{I}}{\argmax}~ \theta^j |} & \text{if $\theta^i=\max(\theta)$} \\ 0 & \text{otherwise} \end{cases} \quad \text{ for $i=1,2, \ldots, I$.}\] But, ties can be broken in other ways. Extending the notation from the screening model, let $\delta_{\theta} \in \Delta(\Theta)$ denote the measure that places probability one on type profile $\theta \in \Theta$.

\subsection{Robust mechanisms}

It is again established that there are few restrictions placed on the designer's mechanism if the only requirement is that she chooses a best response to an adversarial LPS; any mechanism that is ``efficient at the bottom" and does not yield excessive rent is robustly optimal.

\begin{proposition}[Auction: characterization of robust mechanisms]\label{thm_auction_robust}
A mechanism $(Q,P) \in \mathcal{M}$ is robust if and only if
\begin{enumerate}
    \item it is efficient for the lowest type profile, $\underline{\theta}$;
    \item the seller extracts full surplus under the lowest type profile,
    \[\sum^I_{i=1} P^i(\underline{\theta})= \max(\underline{\theta});\]
    \item and the auctioneer obtains the lowest payoff from the lowest type profile,  \[ v((Q,P),\theta) \geq v((Q,P), \underline{\theta}) \quad \text{for all $\theta \in \Theta$}.\]
\end{enumerate}
\end{proposition}

Supplemental Appendix \ref{app_ex_robust_auction} exhibits a robust, inefficient, and weakly dominated mechanism.

\subsection{Perfectly robust mechanisms}

The following result establishes that, if there are two types, a mechanism is perfectly robust if and only if it is efficient and maximal. More generally, efficiency must always arise ``at the top" and ``at the bottom".

\begin{proposition}[Auction: characterization of perfectly robust mechanisms]\label{thm_auction_perfect}
If there are two types ($N=2$), then a mechanism is perfectly robust if and only if it is efficient and maximal. More generally, in any perfectly robust mechanism $(Q,P) \in \mathcal{M}$, $P$ satisfies \eqref{rev_auction} and the allocation rule is efficient for any type profile $\theta \in \Theta$ with $\max(\theta) \in \{\theta_1, \theta_N\}$.
\end{proposition}

Perfectly robust mechanisms again need not be efficient ``in the middle". Supplemental Appendices \ref{app_ex_auction_ineff2} and \ref{app_ex_auction_misallocation} show that inefficiency can arise either because the auctioneer withholds the good, or because she misallocates the good to a bidder that does not have the highest value. These inefficiencies are precisely those that arise in the standard independent private values auction model (see, e.g., \cite{myerson1981optimal} and \cite{ausubel1999optimality}). 

\subsection{Properly robust mechanisms}

The main result for the auction model follows below.

\begin{proposition}[Auction: characterization of properly robust mechanisms]\label{thm_auction_proper}
A mechanism $(Q,P) \in \mathcal{M}$ is properly robust if and only if it is efficient and maximal with ties broken uniformly for any type profile that does not contain the highest type, i.e., for any $\theta \in \Theta$ such that $\max(\theta) < \theta_N$,
\[Q^i(\theta)= \begin{cases} \frac{1}{| \underset{j \in \mathcal{I}}{\argmax}~ \theta^j |} & \text{if $\theta^i=\max(\theta)$}\\ 0 & \text{otherwise.} \end{cases}\]
\end{proposition}

The first part of the proof constructs an LPS $\mu$ that has full support and is strongly adversarial with respect to the mechanisms identified in the statement of Proposition \ref{thm_auction_proper}. The construction is illustrated for the two bidder, three type case in Figure \ref{fig_LPS_auction}. The second part of the proof confirms that \textit{every} efficient and maximal mechanism is $\mu$-optimal. Because $\mu$ is strongly adversarial with respect to the mechanisms identified in Proposition \ref{thm_auction_proper} and those mechanisms are efficient and maximal, they are properly robust. (On the other hand, because $\mu$ is not strongly adversarial with respect to efficient and maximal mechanisms that do not satisfy the restrictions on the tie-breaking rule stated in Proposition \ref{thm_auction_proper}, this argument does \textit{not} establish that they are properly robust.) The third part of the proof shows that any properly robust mechanism must be efficient and maximal with ties broken uniformly whenever the highest type does not appear. 

\begin{figure}
	\centering
	\begin{tikzpicture}[
	every node/.style={font=\small},
	profile/.style={
		circle,
		draw=black,
		text=black,
		semithick,
		minimum size=1.7cm,
		inner sep=1pt,
		align=center
	}
	]
	
	% Colors for the support groups
	\definecolor{c1}{RGB}{228,26,28}  % red
	\definecolor{c2}{RGB}{255,127,0}  % orange
	\definecolor{c3}{RGB}{152,78,163} % purple
	\definecolor{c4}{RGB}{77,175,74}  % green
	\definecolor{c5}{RGB}{55,126,184} % blue
	\definecolor{c6}{RGB}{166,86,40}  % brown
	
	% Type profiles and corresponding beliefs
	\node[profile,fill=c1!20] (t11) at (0,3.6)
	{$(\theta_1,\theta_1)$\\[3pt]$\mu_1$};
	\node[profile,fill=c2!20] (t12) at (2.6,3.6)
	{$(\theta_1,\theta_2)$\\[3pt]$\mu_2$};
	\node[profile,fill=c3!20] (t13) at (5.2,3.6)
	{$(\theta_1,\theta_3)$\\[3pt]$\mu_3$};
	
	\node[profile,fill=c2!20] (t21) at (0,1.8)
	{$(\theta_2,\theta_1)$\\[3pt]$\mu_2$};
	\node[profile,fill=c4!20] (t22) at (2.6,1.8)
	{$(\theta_2,\theta_2)$\\[3pt]$\mu_4$};
	\node[profile,fill=c5!20] (t23) at (5.2,1.8)
	{$(\theta_2,\theta_3)$\\[3pt]$\mu_5$};
	
	\node[profile,fill=c3!20] (t31) at (0,0)
	{$(\theta_3,\theta_1)$\\[3pt]$\mu_3$};
	\node[profile,fill=c5!20] (t32) at (2.6,0)
	{$(\theta_3,\theta_2)$\\[3pt]$\mu_5$};
	\node[profile,fill=c6!20] (t33) at (5.2,0)
	{$(\theta_3,\theta_3)$\\[3pt]$\mu_6$};
	
	\end{tikzpicture}
	\caption{An LPS $\mu=(\mu_1,\mu_2,\mu_3,\mu_4,\mu_5,\mu_6)$
		in which each belief is uniform over its support and the supports are disjoint.
		The label $\mu_k$ inside each circle identifies the belief
		whose support contains that type profile.}
	\label{fig_LPS_auction}
\end{figure}

The intuition behind the result is as follows. By Proposition \ref{proper_leximin}, proper robustness of a mechanism requires lexicographic optimality of its sorted payoff vector across all type profiles. Any deviation from efficiency or the stated tie-breaking restrictions reduces revenue at some type profile in a way that cannot be offset by gains elsewhere. Introducing inefficiency might increase revenue in Bayesian settings by allowing the auctioneer to extract rent from higher types. But profiles containing higher types are of lower lexicographic priority, making inefficiency suboptimal. The intuition behind the optimality of uniform tie-breaking is subtler: if ties at any type profile $\theta \in \Theta$ such that $\max(\theta) < \theta_N$ are broken unevenly, then some agent receives the good with a strictly higher-than-average probability, reducing the maximal rent that can be extracted from him at profiles where his value is slightly higher. To lexicographically maximize payoffs, the auctioneer is therefore better off equalizing allocation probabilities across tied agents. This logic does not extend to type profiles in which the maximum type is $\theta_N$: at such profiles, properness does not restrict how ties are broken.

Observe that the efficient and maximal mechanism with uniform tie-breaking is properly robust. It is implemented by a sealed-bid or ascending clock auction in which the agent with the maximum bid receives the good, with ties broken uniformly at random when there are multiple bidders with the same maximum bid. The winning bidder pays the auctioneer an amount equal to his bid if there are multiple bidders with the same maximum bid. Otherwise, he pays the auctioneer an amount between the second-highest bid $\theta_n$ and the next-highest bidding increment $\theta_{n+1}$,
\[\left(\frac{1}{J+1}\right) \theta_n+ \left(1-\frac{1}{J+1}\right) \theta_{n+1},\]
where $J$ is the number of bidders with the second-highest bid. As the maximal distance between adjacent types converges to zero, the transfer rule converges to that under a second-price auction with uniform tie-breaking (equivalently, that under the Vickrey-Clarke-Groves mechanism with the Clarke pivot rule \citep{Vickrey_1961,Clarke_1971,Groves_1973}).

\subsection{Bayesian foundations}\label{auction_bayes}
 
 Assuming that prices are determined by \eqref{rev_auction}, an allocation rule  is a part of a Bayesian-optimal mechanism with respect to a full support prior $\rho \in \Delta(\Theta)$ if and only if it solves
\begin{equation}
\begin{aligned}
&\max_{Q: \Theta \rightarrow [0,1]^I} \quad \sum^I_{i=1} \sum^N_{j=1} \sum_{ \theta^{-i} \in \Theta^{-i}} \rho(\theta_j, \theta^{-i}) Q^i(\theta_j, \theta^{-i}) v^i(\theta_j, \theta^{-i})\\
&\text{subject to}\\
& v^i(\theta_j, \theta^{-i})=  \theta_j-(\theta_{j+1}-\theta_j)\frac{\sum^N_{k=j+1} \rho(\theta_k,\theta^{-i})}{\rho(\theta_j, \theta^{-i})} \quad \text{for $j<N$}\\
&v^i(\theta_N, \theta^{-i})=\theta_N\\
& Q^i(\cdot,\theta^{-i})~\text{increasing for all $i$ and $\theta^{-i} \in \Theta^{-i}$}\\
& \sum^I_{i=1} Q^i(\theta) \leq 1~\text{for all $i$ and $\theta \in \Theta$,}
\end{aligned} \notag \label{auction_Bayesian}
\end{equation}
where $v^i(\theta)$ is the ``virtual value" of bidder $i$ under type profile $\theta$. Ignoring the monotonicity constraint on $Q^i(\cdot,\theta^{-i})$, it is immediate that any $\rho$-optimal mechanism allocates the good with probability one to bidder $i$ if $v^i(\theta)>\max\{0,\max_{j \neq i} v^j(\theta)\}$. If multiple bidders have the same greatest strictly-positive virtual value, then the good is allocated with probability one and ties can be broken arbitrarily.

Suppose there are two agents ($I=2$) and three types ($N=3$). Consider the LPS $\mu=(\mu_1,\mu_2,\mu_3,\mu_4,\mu_5,\mu_6)$ depicted in Figure \ref{fig_LPS_auction} against which every properly robust mechanism is $\mu$-optimal. As observed in Proposition \ref{lex_Bayes}, there is a precise sense in which properly robust mechanisms outperform others in the limit of particular sequences of Bayesian priors converging to $\delta_{\underline{\theta}}$. Consider a sequence $(r_\ell)_\ell \in ((0,1)^{5})^{\mathbb{N}}$ for which $(r^1_\ell, r^2_\ell, r^3_\ell, r^4_\ell, r^5_\ell) \rightarrow (0,0,0,0,0)$. The corresponding Bayesian prior $\rho_\ell:=(r_\ell \square \mu)$ is displayed below:
\begin{equation}
\setlength{\arraycolsep}{1pt}
\begin{array}{rcl rcl rcl}
\rho_\ell(\theta_1,\theta_1) & = & 1 - r^1_\ell
&
\rho_\ell(\theta_2,\theta_1) & = & \frac{1}{2}\, r^1_\ell (1 - r^2_\ell)
&
\rho_\ell(\theta_3,\theta_1) & = & \frac{1}{2}\, r^1_\ell r^2_\ell (1 - r^3_\ell)
\\
\rho_\ell(\theta_1,\theta_2) & = & \frac{1}{2}\, r^1_\ell (1 - r^2_\ell)
&
\rho_\ell(\theta_2,\theta_2) & = & r^1_\ell r^2_\ell r^3_\ell (1 - r^4_\ell)
&
\rho_\ell(\theta_3,\theta_2) & = & \frac{1}{2}\, r^1_\ell r^2_\ell r^3_\ell r^4_\ell (1 - r^5_\ell)
\\
\rho_\ell(\theta_1,\theta_3) & = & \frac{1}{2}\, r^1_\ell r^2_\ell (1 - r^3_\ell)
&
\rho_\ell(\theta_2,\theta_3) & = & \frac{1}{2}\, r^1_\ell r^2_\ell r^3_\ell r^4_\ell (1 - r^5_\ell)
&
\rho_\ell(\theta_3,\theta_3) & = & r^1_\ell r^2_\ell r^3_\ell r^4_\ell r^5_\ell.
\end{array}\notag
\end{equation}
The virtual values for $i=1,2$ are then 
\begin{equation}
\begin{array}{rcl rcl}
v^i_\ell(\theta_1,\theta_1)
&=&
\theta_1-(\theta_2-\theta_1)
\left(
\frac{ \frac{1}{2} r^1_\ell\!\left((1-r^2_\ell)+r^2_\ell(1-r^3_\ell)\right) }
     {(1-r^1_\ell)}
\right)
&
v^i_\ell(\theta_2,\theta_1)
&=&
\theta_2-(\theta_3-\theta_2)
\left(
\frac{ r^2_\ell(1-r^3_\ell) }{1-r^2_\ell}
\right)
\\[1.5mm]

v^i_\ell(\theta_1,\theta_2)
&=&
\theta_1-(\theta_2-\theta_1)
\left(
\frac{ r^2_\ell r^3_\ell\!\left((1-r^4_\ell)+\tfrac12 r^4_\ell(1-r^5_\ell)\right) }
     {\tfrac12(1-r^2_\ell)}
\right)
&
v^i_\ell(\theta_2,\theta_2)
&=&
\theta_2-(\theta_3-\theta_2)
\left(
\frac{ \tfrac12 r^4_\ell(1-r^5_\ell) }{1-r^4_\ell}
\right)
\\[1.5mm]

v^i_\ell(\theta_1,\theta_3)
&=&
\theta_1-(\theta_2-\theta_1)
\left(
\frac{  r^3_\ell r^4_\ell\!\left(\frac{1}{2} (1-r^5_\ell)+r^5_\ell\right) }
     {\frac{1}{2}(1-r^3_\ell)}
\right)
&
v^i_\ell(\theta_2,\theta_3)
&=&
\theta_2-(\theta_3-\theta_2)
\left(
\frac{ r^5_\ell }{\frac{1}{2} (1-r^5_\ell)}
\right)
\end{array}\notag
\end{equation}
\[v^i_\ell(\theta_3,\theta_3)=v^i_\ell(\theta_3,\theta_2)=v^i_\ell(\theta_3,\theta_1)=\theta_3.\] 
Note that, as $\ell \rightarrow \infty$, $v^i_\ell(\theta^i,\theta^{-i})$ converges to $\theta^i$. That is, if lower type profiles are much more likely than higher type profiles, then virtual values converge to actual values and distortions become unprofitable. The role of proper robustness in refining the tie-breaking rule cannot be demonstrated via such an example --- against the justifying LPS constructed in the proof of Proposition \ref{thm_auction_proper}, any efficient and maximal mechanism is $\mu$-optimal. However, as previously remarked, $\mu$ is not strongly adversarial with respect to an efficient and maximal mechanism that does not satisfy the tie-breaking restrictions stated in Proposition \ref{thm_auction_proper}.

\section{Public good provision}\label{sec_pub} 

This section departs from the auction model in Section \ref{sec_auction} by assuming that the good to be allocated is public. The model differs from those in the classic mechanism design literature on public good provision (e.g., \cite{Clarke_1971} and \cite{Groves_1973}) due to the designer's objective to maximize profit rather than efficiency. Nevertheless, the profit-maximization objective facilitates comparison to the analysis of the previous sections.  It is shown that robust mechanisms are typically inefficient, and that robustness and perfect robustness generally permit a large set of mechanisms. Nevertheless, proper robustness pins down a unique mechanism. An asymptotic exercise in the spirit of \cite{rob1989pollution} and \cite{mailath1990asymmetric} establishes that, if the economy grows large and the per-capita cost of the public good remains constant, a stark form of inefficiency arises.

\subsection{Environment}

There is a single profit-maximizing firm that can produce a public good consumed by $I \geq 2$ agents $i \in \{1,2, \ldots, I\}=: \mathcal{I}$. Each agent $i$ has private information about their payoff from the production of the public good $\theta^i \in \Theta^i:= \{\theta_\ell, \theta_h\}$, where $\theta_h>0$ and $\theta_h>\theta_\ell$ (note that $\theta_\ell$ need not be positive). Let the set of type profiles be denoted by $\Theta:= \prod^I_{i=1} \Theta^i$, let $\underline{\theta}:=(\theta_\ell, \ldots, \theta_\ell)$, and let $\overline{\theta}:= (\theta_h, \ldots, \theta_h)$. The ex post payoff of agent $i$ with type $\theta^i$ is given by
\[q \theta^i-p^i, \] where $q \in [0,1]$ is the probability with which the public good is produced and $p^i \in \mathbb{R}$ is the transfer from agent $i$ to the firm. The firm's ex post payoff is given by
\[\left(\sum^I_{i=1} p^i \right)-q c, \]
where $c := \gamma I$ is the total cost of producing the public good and $\gamma \in (\theta_\ell, \theta_h)$ is the per-capita cost.  

The firm has commitment power. By the Revelation Principle, without loss of optimality, it can restrict attention to direct mechanisms.  A \textbf{(direct) mechanism} is an allocation rule and transfer rule,
\begin{equation}
Q: \Theta \rightarrow [0,1] \quad \text{and} \quad P: \Theta \rightarrow \mathbb{R}^I,\notag
\end{equation}
that together are \textbf{(dominant strategy) incentive compatible} and \textbf{(ex post) individually rational}: for all $ \theta \in \Theta$ and all $i=1,2, \ldots, I$, 
\[ Q(\theta^i, \theta^{-i}) \theta^i-P^i(\theta^i, \theta^{-i}) \geq  Q(\hat{\theta}^i, \theta^{-i}) \theta^i-P^i(\hat{\theta}^i, \theta^{-i}) \quad \text{for $\hat{\theta}^i \neq \theta^i$} \]
and
\[ Q(\theta^i, \theta^{-i}) \theta^i-P^i(\theta^i, \theta^{-i}) \geq 0.\] Let $\mathcal{M}$ denote the set of all mechanisms. Given a mechanism $(Q,P) \in \mathcal{M}$ and a type profile $\theta \in \Theta$, the firm obtains a payoff of
\[v((Q,P), \theta)= \left(\sum^I_{i=1} P^i(\theta) \right)- Q(\theta)c.\] As in the previous environments, it is easy to show that randomization over mechanisms results in a mechanism that is payoff equivalent to one in $\mathcal{M}$; henceforth, without loss of generality, the firm chooses an element of $\mathcal{M}$. 

The allocation rule $Q: \Theta \rightarrow [0,1]$ is \textbf{efficient for type profile $\theta$} if $\sum^I_{i=1} \theta^i > c$ and $Q(\theta)=1$, $\sum^I_{i=1} \theta^i < c$ and $Q(\theta)=0$, or $\sum^I_{i=1} \theta^i=c$. The allocation rule $Q: \Theta \rightarrow [0,1]$ is \textbf{efficient} if it is efficient for all type profiles.  The mechanism $(Q,P) \in \mathcal{M}$ is \textbf{efficient and maximal} if $Q$ is the efficient allocation rule for which $Q(\theta)=0$ if $\sum^I_{i=1} \theta^i=c$
and $P$ satisfies, for all $i=1,2, \ldots, I$ and all $\theta^{-i} \in \Theta^{-i}$,
\begin{equation}
P^i(\theta_\ell, \theta^{-i}) = Q(\theta_\ell, \theta^{-i}) \theta_\ell \quad \text{and} \quad P^i(\theta_h, \theta^{-i}) = Q(\theta_\ell, \theta^{-i}) \theta_\ell+ (Q(\theta_h,\theta^{-i})-Q(\theta_\ell,\theta^{-i})) \theta_h. \label{rev_pub}
\end{equation}
Again, let $\delta_{\theta} \in \Delta(\Theta)$ denote the measure that places probability one on type profile $\theta$.

The following notation will be useful. Let \[S_k:= I \theta_\ell + k (\theta_h-\theta_\ell)-c\] be the surplus generated when the public good is produced in any type profile in which $k$ agents have type $\theta_h$. Let $k^* :=\min\{k \in \{1, \ldots, I\}: S_k>0\}$ be the lowest number of agents with valuation $\theta_h$ such that strictly positive surplus is generated from producing the public good. Note that, because $c=\gamma I$, it is efficient to produce the public good whenever the fraction of agents with value $\theta_h$ is at least a threshold $\rho \in (0,1)$ that does not depend on the number of agents:
\[ S_k= I \theta_\ell + k (\theta_h-\theta_\ell)- \gamma I \geq 0 \iff \frac{k}{I} \geq  \frac{\gamma-\theta_\ell}{\theta_h-\theta_\ell}=: \rho.  \]

\subsection{Robust mechanisms}
The following proposition characterizes the set of robust mechanisms.

\begin{proposition}[Public good: characterization of robust mechanisms]\label{thm_pub_robust}
A mechanism $(Q,P) \in \mathcal{M}$ is robust if and only if the firm's ex post payoff is always at least zero: for all $\theta \in \Theta$,
\[v((Q,P),\theta)=\left(\sum^I_{i=1} P^i(\theta) \right)-Q(\theta) c \geq 0. \]
\end{proposition}

It is again easy to identify robust, but weakly dominated, mechanisms.

\begin{example}\label{ex_public_robust}
	Consider the mechanism $(Q,P) \in \mathcal{M}$ in which the public good is never produced: $Q(\theta)=P^i(\theta)=0$ for all $\theta \in \Theta$ and all $i$. By Proposition \ref{thm_pub_robust}, it is robust because it yields the firm a payoff of zero under every type profile. However, it is weakly dominated by the mechanism $(\hat{Q},\hat{P}) \in \mathcal{M}$ in which $\hat{Q}(\theta)=1$ if and only if $\theta=\overline{\theta}$ and $\hat{P}$ satisfies \eqref{rev_pub} with $\hat{Q}$ replacing $Q$. In particular, $(\hat{Q},\hat{P})$ yields a payoff of zero for all $\theta \neq \overline{\theta}$ and extracts full surplus $I \theta_h-c>0$  under the profile $\theta=\overline{\theta}$.
\end{example}

An immediate corollary of Proposition \ref{thm_pub_robust} is that the efficient and maximal mechanism is robust if and only if $I \theta_h >c \geq (I-1) \theta_h + \theta_\ell$ (recall the maintained assumption that $c \in (I \theta_\ell, I \theta_h)$). To see why, consider the efficient allocation rule in which the public good is provided if and only if all agents have the maximum valuation, i.e., $\theta=\overline{\theta}$. If transfers are determined by \eqref{rev_pub}, then the firm obtains a payoff of $I \theta_h -c > 0$ under $\overline{\theta}$. Under all other type profiles, the firm obtains a payoff of zero. So, the firm always obtains at least a payoff of zero and Proposition \ref{thm_pub_robust} implies that the mechanism is robust. On the other hand, if $c<(I-1) \theta_h+\theta_\ell$, then under the type profile $\theta=\overline{\theta}$, the public good is produced independently of an individual agent's report in any efficient allocation rule. Hence, to preserve incentive compatibility, no agent can be charged a price strictly larger than $\theta_\ell$ for reporting their value truthfully. The firm thus obtains a payoff no higher than $I \theta_\ell-c<0$. This result is stated below for convenience.

\begin{corollary}[(Sub-)optimality of efficient and maximal mechanism]\label{corr_pub_ineff}
The efficient and maximal mechanism is robust if and only if $c \in [(I-1) \theta_h+\theta_\ell, I \theta_h)$.
\end{corollary}

Corollary \ref{corr_pub_ineff} isolates the economic force behind inefficiency in the model: under the incentive compatibility and individual rationality constraints, efficient provision need not permit sufficient revenue extraction by the designer ``at the top".

\subsection{Perfectly robust mechanisms}

By Proposition \ref{perfect_admissibility}, requiring the firm to choose a perfectly robust mechanism rules out selection of a weakly dominated mechanism. Hence, the weakly dominated mechanism constructed in Example \ref{ex_public_robust} cannot be perfectly robust. The following proposition establishes that, when $c \in [(I-1)\theta_h+\theta_\ell, I\theta_h)$, the efficient and maximal mechanism is the unique perfectly robust mechanism. More generally, efficiency must always arise ``at the top" and ``at the bottom".

\begin{proposition}[Public good: characterization of perfectly robust mechanisms]\label{thm_pub_perfect}
If $c \in [(I-1) \theta_h+\theta_\ell, I \theta_h)$, then the efficient and maximal mechanism is the unique perfectly robust mechanism. More generally, in any perfectly robust mechanism $(Q,P) \in \mathcal{M}$, the allocation rule $Q$ is efficient for any type profile $\theta \in \Theta$ such that $\sum^I_{i=1} \theta^i \leq c$ or $\theta=\overline{\theta}$, and the transfer rule $P$ satisfies \eqref{rev_pub}.
\end{proposition}

Supplemental Appendix \ref{app_ex_pub} demonstrates that, outside the case in which $c \in [(I-1) \theta_h+\theta_\ell, I \theta_h)$, perfection need not impose much discipline on the set of selected mechanisms.

\subsection{Properly robust mechanisms}

 The following proposition identifies the unique properly robust mechanism. In the statement of the proposition and throughout, the empty product equals one by convention. 

\begin{proposition}[Public good: characterization of properly robust mechanisms]\label{thm_pub_proper}
The unique properly robust mechanism is the mechanism $(Q,P) \in \mathcal{M}$ for which
\[ Q(\theta)= \begin{cases} \frac{q_K}{q_I} & \text{if $|\{i: \theta^i=\theta_h\}|=K$ and $\sum^I_{i=1} \theta^i>c$}\\ 0 & \text{otherwise} \end{cases},\] where
\[q_K := \sum^K_{k=k^*} \left(\frac{1}{S_k} \prod^K_{j=k+1} \frac{j (\theta_h-\theta_\ell)}{S_j} \right) \quad \text{for $K \in \{k^*, \ldots, I\}$,}\]
and $P$ satisfies \eqref{rev_pub}.
\end{proposition}

Inspecting the statement of Proposition \ref{thm_pub_proper} reveals that the properly robust mechanism is inefficient whenever $c < (I-1) \theta_h+\theta_\ell$. In particular, the probability with which the public good is produced is strictly smaller than one when $\sum^I_{i=1} \theta^i=(I-1) \theta_h + \theta_\ell$ even though the sum of valuations strictly exceeds the cost of production. The allocation rule is depicted for a particular parameterization in Figure \ref{fig_converge} and its asymptotic properties are explored in Section \ref{asymptotic}. Supplemental Appendix \ref{app_fig_converge} contains the calculated values for the Figure.

The proof of Proposition \ref{thm_pub_proper} proceeds by first showing that any properly robust mechanism is anonymous, i.e., depends only on the number of agents with type $\theta_h$. Necessity of the conditions defining the mechanism in the statement of Proposition \ref{thm_pub_proper} is then established by showing that leximin optimality requires equalizing the firm's payoff across type profiles in which strictly positive surplus is generated. By Proposition \ref{proper_leximin}, it follows that these conditions are also necessary for proper robustness. Sufficiency is established by constructing a strongly adversarial and full support LPS with two beliefs against which the mechanism is optimal. The first belief, $\mu_1$, has support equal to the set of type profiles that do not generate strictly positive surplus. The second belief, $\mu_2$, has support equal to the set of type profiles generating strictly positive surplus. This construction yields an immediate Bayesian foundation for the unique properly robust mechanism --- it is optimal against any strict convex combination of $\mu_1$ and $\mu_2$. Other mechanisms can also be optimal against this convex combination. But, as in the auction model of Section \ref{sec_auction}, $\mu$ is not strongly adversarial with respect to these mechanisms.

\subsection{Asymptotic inefficiency}\label{asymptotic}

The unique properly robust allocation rule in an economy with $I$ agents is characterized by an $(I+1)$-vector of probabilities $(Q^I_{k})_{k}$, where $Q^I_k$ is the probability with which the public good is produced under $Q^I$ given any type profile in which there are exactly $k$ agents with type $\theta_h$. This $(I+1)$-vector can be summarized by the function $f^I :[0,1] \rightarrow [0,1]$ defined by
\[f^I(x)=Q^I_{\lfloor xI \rfloor},\] where $\lfloor xI \rfloor$ denotes the greatest integer less than or equal to $xI$. That is, $f^I(x)$ denotes the probability with which the good is produced when there are $k=\lfloor x I \rfloor$ agents with type $\theta_h$. Equivalently, $f^I(x)$ is the maximum probability with which the public good is produced when the fraction of agents with type $\theta_h$ is at most $x \in [0,1]$. 

From Proposition \ref{thm_pub_proper}, for all $I$, $f^I(1)=1$ and $f^I(x)=0$ if $x \leq \rho$, where $\rho$ is the efficient threshold. The following proposition establishes further that, for any $x \in (\rho, 1)$, the probability of provision when at most a fraction $x$ of agents have type $\theta_h$ goes to zero as the number of agents grows large. Moreover, for any $\varepsilon>0$, convergence is uniform on the interval $[0,1-\varepsilon]$ at an exponential rate.

\begin{figure}
	\centering
	\begin{tikzpicture}
	\begin{axis}[
	width=13cm,
	height=8cm,
	xmin=0, xmax=1,
	ymin=0, ymax=1,
	xlabel={Fraction of $\theta_h$ agents $x=\frac{k}{I}$},
	ylabel={Probability of provision},
	axis lines=left,
	tick style={black},
	legend style={
		draw=none,
		at={(0.02,0.98)},
		anchor=north west,
		cells={align=left}
	}
	]
	
	% Computed with exact rational arithmetic; rounded to five significant digits.
	% Efficient cutoff rho = 0.5
	\addplot[gray!60, thick] coordinates {(0.5,0) (0.5,1)};
	\addlegendentry{Efficient threshold $\rho$}
	
	% I = 5
	\addplot[
	only marks,
	mark=*,
	mark size=3.5pt,
	color={rgb,255:red,0; green,114; blue,178}
	] coordinates {
		(0,0)
		(0.2,0)
		(0.4,0)
		(0.6,0.16129)
		(0.8,0.48387)
		(1,1)
	};
	\addlegendentry{$I=5$}
	
	% I = 15
	\addplot[
	only marks,
	mark=square*,
	mark size=2.5pt,
	color={rgb,255:red,213; green,94; blue,0}
	] coordinates {
		(0,0)
		(0.066667,0)
		(0.13333,0)
		(0.2,0)
		(0.26667,0)
		(0.33333,0)
		(0.4,0)
		(0.46667,0)
		(0.53333,0.00045778)
		(0.6,0.0028993)
		(0.66667,0.011689)
		(0.73333,0.036801)
		(0.8,0.098187)
		(0.86667,0.23212)
		(0.93333,0.49998)
		(1,1)
	};
	\addlegendentry{$I=15$}
	
	% I = 55
	\addplot[
	only marks,
	mark=triangle*,
	mark size=3.0pt,
	color={rgb,255:red,0; green,158; blue,115}
	] coordinates {
		(0,0)
		(0.018182,0)
		(0.036364,0)
		(0.054545,0)
		(0.072727,0)
		(0.090909,0)
		(0.10909,0)
		(0.12727,0)
		(0.14545,0)
		(0.16364,0)
		(0.18182,0)
		(0.2,0)
		(0.21818,0)
		(0.23636,0)
		(0.25455,0)
		(0.27273,0)
		(0.29091,0)
		(0.30909,0)
		(0.32727,0)
		(0.34545,0)
		(0.36364,0)
		(0.38182,0)
		(0.4,0)
		(0.41818,0)
		(0.43636,0)
		(0.45455,0)
		(0.47273,0)
		(0.49091,0)
		(0.50909,1.5266e-15)
		(0.52727,3.0022e-14)
		(0.54545,3.6057e-13)
		(0.56364,3.1939e-12)
		(0.58182,2.2712e-11)
		(0.6,1.3627e-10)
		(0.61818,7.1281e-10)
		(0.63636,3.3264e-9)
		(0.65455,1.4088e-8)
		(0.67273,5.4871e-8)
		(0.69091,1.9858e-7)
		(0.70909,6.7345e-7)
		(0.72727,0.0000021550)
		(0.74545,0.0000065449)
		(0.76364,0.000018958)
		(0.78182,0.000052592)
		(0.8,0.00014025)
		(0.81818,0.00036063)
		(0.83636,0.00089671)
		(0.85455,0.0021613)
		(0.87273,0.0050606)
		(0.89091,0.011533)
		(0.90909,0.025630)
		(0.92727,0.055622)
		(0.94545,0.11806)
		(0.96364,0.24537)
		(0.98182,0.50000)
		(1,1)
	};
	\addlegendentry{$I=55$}
	
	\end{axis}
	\end{tikzpicture}
	\caption{Asymptotic inefficiency under properly robust mechanism ($\theta_h=1$, $\theta_\ell=0$, and $\gamma=\frac{1}{2}$).}\label{fig_converge}
\end{figure}

\begin{proposition}[Public good: asymptotic inefficiency of properly robust mechanism]\label{thm_pub_asymptotic}
For any threshold $x \in [0,1)$, the probability with which the public good is produced in the unique properly robust allocation rule when the fraction of agents with type $\theta_h$ is at most $x$ vanishes as the number of agents grows large. That is, $f^I: [0,1] \rightarrow [0,1]$ converges pointwise to the function $f^\infty: [0,1] \rightarrow [0,1]$ defined by
\[ f^\infty(x)= \begin{cases} 0 & \text{if $x \in [0,1)$}\\
1 & \text{if $x=1$} \end{cases}.\]
Moreover, for any $\varepsilon \in (0,1)$, convergence is uniform on $[0,1-\varepsilon]$ at exponential rate $\lambda_\varepsilon:= \varepsilon(-\ln(1-\rho))$: for all $I$,
\[ \sup_{x \in [0,1-\varepsilon]} |f^I(x)-f^\infty(x)| \leq \exp(-\lambda_\varepsilon I).\]
\end{proposition}

For any fixed fraction of agents with type $\theta_h$ in the interval $(\rho,1)$, providing the public good with probability one is efficient. Proposition \ref{thm_pub_asymptotic} shows that, at or below any such fraction, the probability of provision under the properly robust mechanism vanishes as the number of agents grows large. A comparison to the literature is in order. \cite{rob1989pollution} considers a Bayesian setting in which a profit-maximizing firm decides whether or not to build a pollution-generating plant, and compensates residents for pollution damages. Each resident holds private information about their individual damage and has the power to veto the construction of the plant. So, the firm chooses a mechanism subject to (Bayesian) incentive compatibility and individual rationality constraints. \cite{mailath1990asymmetric} consider the entire set of mechanisms that satisfy Bayesian incentive compatibility, individual rationality, and budget balance constraints (no objective function for the designer is specified). Both \cite{rob1989pollution} and \cite{mailath1990asymmetric} identify conditions on the per-capita cost of providing the public good and on the distribution over type profiles under which the probability of public good provision approaches zero as the number of agents grows large.  In contrast, this section considers the robust design of profit-maximizing mechanisms under dominant strategy incentive compatibility and ex post individual rationality constraints. Given the non-Bayesian setting, the inefficiency result is stated in ex post terms. Nevertheless, a straightforward consequence of the proof of Proposition \ref{thm_pub_proper} is that, under the sequence of Bayesian priors that rationalize the corresponding sequence of properly robust mechanisms, the probability of provision also approaches zero.

\section{Discussion and related literature}\label{discuss}

This paper contributes a framework for robust mechanism design and a collection of application-specific results. The strongest proposed refinement, proper robustness, provides a belief-based characterization of the leximin criterion within an LPS framework that also nests the admissibility refinement of the maxmin criterion. In the applications, proper robustness has sharp efficiency implications: in the screening and auction models studied, it selects efficient and maximal mechanisms. In the public good model, the unique properly robust mechanism may be inefficient. These inefficiencies become severe as the economy grows large when the per-capita cost of providing the public good is fixed. The key economic driver of (in)efficiency is the relationship between allocation monotonicity and payoff monotonicity for the designer under the efficient allocation rule.

An alternative approach to sharpening maxmin predictions in mechanism design problems is to impose constraints on the designer's uncertainty set. For instance, \cite{bergemann2011robust} and 
\cite{Carrasco_JET2018} study the problem of selling a single indivisible good to a buyer of unknown type. \cite{bergemann2011robust} assume that the seller has knowledge that the true distribution lies in some neighborhood of a baseline distribution, whereas \cite{Carrasco_JET2018} assume that the seller knows the first $N$ moments of the distribution. The main result in \cite{bergemann2011robust} is that the seller chooses a posted price to maximize her Bayesian payoff against a worst-case distribution. \citet{Carrasco_JET2018} show that the seller's worst-case payoff is a linear combination of the known moments of the distribution over types. Proposition \ref{thm_robust} of this paper is consistent with \cite{bergemann2011robust}'s result; if any distribution is possible, i.e., the seller considers an arbitrarily large neighborhood around the known distribution, then the seller maximizes her payoff against the point mass on the lowest buyer valuation type. \cite{bergemann2011robust} and \cite{Carrasco_JET2018} impose additional assumptions on the seller's knowledge beyond the set of feasible types to provide sharper predictions. This paper offers a complementary modeling approach.

Within the broader literature on robust mechanism design, the point of view taken is inspired by \cite{borgers2017no}. \cite{borgers2017no} critiques the maxmin foundations for dominant strategy mechanisms provided by \cite{ChungEly_2007} by showing that such mechanisms are weakly dominated.\footnote{\cite{borgers2025undominated} characterize mechanisms that are undominated without imposing maxmin selection of mechanisms. \cite{mishra2025undominated} characterize undominated mechanisms in a monopoly regulation setting and use their characterization to identify maxmin optimal mechanisms.} Others have built upon this implicitly lexicographic approach to refining the maxmin criterion: \cite{dworczak2022preparing} assume that an information designer first optimizes against a worst-case distribution. Among the set of worst-case optimal mechanisms, the designer then chooses a mechanism that is optimal with respect to a conjecture that is treated as a primitive of the model. \cite{mishra2025robust} apply the two-step approach to procurement with a continuum of cost types. When demand is known, their characterized mechanism is efficient at both endpoints of the type space, but can exhibit interior distortions. This parallels the efficiency implications of perfect robustness in the finite-type screening analysis in this paper. Proper robustness, on the other hand, always selects the efficient and maximal mechanism.
\appendix

\section{Proofs}\label{app_proofs}

\subsection{Proof of Proposition \ref{robust_maxmin}}\label{proof_robust_maxmin}

If a mechanism $\sigma \in \Delta(\mathcal{M})$ is robust, then there exists an LPS $\mu$ that is adversarial with respect to $\sigma$ against which $\sigma$ is $\mu$-optimal. Because $\mu_1(\theta)>0$ only if $\theta \in \argmin_{\theta' \in \Theta} V(\sigma, \theta')$, the first-order payoff of the principal under $\sigma$ is $\min_{\theta' \in \Theta} V(\sigma, \theta')$. Moreover, because $\sigma$ is $\mu$-optimal, no other mechanism $\sigma' \in \Delta(\mathcal{M})$ can obtain a higher first-order payoff than $\min_{\theta' \in \Theta} V(\sigma,\theta')$. Because $\min_{\theta' \in \Theta} V(\sigma',\theta')$ is no larger than the first-order payoff of $\sigma'$ under $\mu$, it follows that
\[ \min_{\theta' \in \Theta} V(\sigma',\theta') \leq \min_{\theta' \in \Theta} V(\sigma,\theta').\] Since $\sigma'$ was chosen arbitrarily, it follows that no other mechanism can obtain a higher minimum payoff, i.e., $\sigma$ is maxmin optimal.

Suppose $\sigma \in \Delta(\mathcal{M})$ is maxmin optimal. Then, 
\[    \bar{v}:= \max_{\sigma' \in \Delta(\mathcal{M})}  \min_{\nu' \in \Delta(\Theta)} \sum_{\theta' \in \Theta} \nu'(\theta') V(\sigma', \theta')\]
is well-defined and, by Theorem 4.2' of \cite{sion1958general}, 
\begin{equation}
   \bar{v} \geq \min_{\nu' \in \Delta(\Theta)} \sup_{\sigma' \in \Delta(\mathcal{M})}  \sum_{\theta' \in \Theta} \nu'(\theta') V(\sigma', \theta') . \label{minimax}
\end{equation}
Now, let $\nu \in \argmin_{\nu' \in \Delta(\Theta)} \sup_{\sigma' \in \Delta(\mathcal{M})}  \sum_{\theta' \in \Theta} \nu'(\theta') V(\sigma', \theta') $ and consider the LPS $\mu=(\nu)$. Observe that for any $\sigma' \in \Delta(\mathcal{M})$, the first-order payoff under $\mu$ is no larger than $\bar{v}$ by \eqref{minimax}. Because $\sigma$ yields at least $\bar{v}$ against any distribution, it follows that $\sigma$ is $\mu$-optimal. Moreover, $\mu$ is adversarial with respect to $\sigma$; $\bar{v}$ is the lowest payoff that $\sigma$ yields across all beliefs. Hence, $\sigma$ is robust.

\subsection{Proof of Proposition \ref{perfect_admissibility}}\label{proof_perfect_admissibility}

It is first proven that if a mechanism $\sigma \in \Delta(\mathcal{M})$ is perfectly robust, then it is maxmin optimal and admissible. Suppose $\sigma \in \Delta(\mathcal{M})$ is perfectly robust. Then, by definition, there exists an LPS, $\mu$, that is adversarial with respect to $\sigma$ and such that $\sigma$ is $\mu$-optimal. Hence, it is robust. By Proposition \ref{robust_maxmin}, it is thus maxmin optimal. To prove that such a mechanism is admissible, observe that if $\sigma \in \Delta(\mathcal{M})$ is not admissible, then there exists a mechanism $\sigma' \in \Delta(\mathcal{M})$ that weakly dominates it. Hence, $\sigma'$ is lexicographically preferred to $\sigma$ given any full support LPS $\mu$. It follows that $\sigma$ cannot be a best response to any full support LPS and, hence, cannot be perfectly robust. 

For the converse, suppose that $|\mathcal{M}|<\infty$ and $\sigma \in \Delta(\mathcal{M})$ is maxmin optimal. Then, by the same argument in the second paragraph of the proof of Proposition \ref{robust_maxmin}, there exists a $\nu \in \Delta(\Theta)$ against which $\sigma$ is optimal that has the property that $\nu(\theta)>0$ only if $V(\sigma, \theta) \leq V(\sigma,\theta')$ for all $\theta' \in \Theta$. If, in addition, $\sigma \in \Delta(\mathcal{M})$ is admissible, then, by Theorem 5.2.5 of \cite{blackwell1954theory}, there exists a full support measure $\rho \in \Delta(\Theta)$ against which $\sigma$ is a best response. Thus, $\sigma \in \Delta(\mathcal{M})$ is $\mu$-optimal for $\mu=(\nu, \rho)$. Because $\mu$ has full support and is adversarial with respect to $\sigma$, it follows that $\sigma$ is perfectly robust.

\subsection{Proof of Proposition \ref{proper_leximin}}\label{proof_proper_leximin}

Throughout, let $\Pi_{(k)}(\sigma)$ denote the $k$-th lowest value in the set $\{V(\sigma, \theta): \theta \in \Theta\}$ and let $\Theta_{(k)}(\sigma):=\{\theta \in \Theta: V(\sigma, \theta)= \Pi_{(k)}(\sigma)\}$ be the set of type profiles yielding the $k$-th lowest value.

First, suppose $\sigma \in \Delta(\mathcal{M})$ is properly robust. Then, there is an LPS $\mu=(\mu_1, \ldots, \mu_K)$ that has full support and is strongly adversarial with respect to $\sigma$, and for which $\sigma$ is $\mu$-optimal. Let $J:=|\{V(\sigma, \theta): \theta \in \Theta\}|$. For each $\ell= 1, \ldots, J$, define $k_\ell:=\min\{k: \exists \theta \in \Theta_{(\ell)}~ \text{such that}~ \mu_k(\theta)>0 \}$. Under any full support and strongly adversarial LPS, $ k_\ell$ is strictly increasing in $\ell$ and, for any index $k < k_\ell$, the support of $\mu_k$ is a subset of $\Theta_{(1)}(\sigma) \cup \cdots \cup \Theta_{(\ell-1)}(\sigma)$. If $J=1$, then $V(\sigma, \theta)=\Pi_{(1)}(\sigma)$ for all $\theta \in \Theta$. Because $\sigma$ is $\mu$-optimal and $\mu$ has full support, it follows that, for any $\sigma' \in \Delta(\mathcal{M})$, either $V(\sigma,\theta)=V(\sigma',\theta)$ for all $\theta \in \Theta$ or there exists a $\theta \in \Theta$ such that $V(\sigma',\theta)<\Pi_{(1)}(\sigma)=V(\sigma, \theta)$. In either case,  $\left( V_{(i)}(\sigma) \right)^N_{i=1} \geq_{L} \left( V_{(i)}(\sigma') \right)^N_{i=1}$. If $J>1$, then the principal's expected payoff under $\mu_k$ for any $k=1, \ldots, k_2-1$ is $\Pi_{(1)}(\sigma)$. Because $\sigma$ is $\mu$-optimal, it follows that, for any $\sigma' \in \Delta(\mathcal{M})$, either $V(\sigma,\theta)=V(\sigma',\theta)$ for all $\theta \in \Theta_{(1)}(\sigma)$ or there exists a $\theta \in \Theta_{(1)}(\sigma)$ such that $V(\sigma',\theta)<\Pi_{(1)}(\sigma)=V(\sigma, \theta)$. In the latter case, $\left( V_{(i)}(\sigma) \right)^N_{i=1} >_{L} \left( V_{(i)}(\sigma') \right)^N_{i=1}$. So, only mechanisms $\sigma' \in \Delta(\mathcal{M})$ satisfying $V(\sigma,\theta)=V(\sigma',\theta)$ for all $\theta \in \Theta_{(1)}(\sigma)$ could possibly be leximin preferred to $\sigma \in \Delta(\mathcal{M})$. Suppose $\sigma' \in \Delta(\mathcal{M})$ is a mechanism such that $V(\sigma,\theta)=V(\sigma',\theta)$ for all $\theta \in \Theta_{(1)}(\sigma) \cup \cdots \cup \Theta_{(j-1)}(\sigma)$. It is shown that either $V(\sigma,\theta)=V(\sigma',\theta)$ for all $\theta \in \Theta_{(1)}(\sigma) \cup \cdots \cup \Theta_{(j)}(\sigma)$ or $\left( V_{(i)}(\sigma) \right)^N_{i=1} >_{L} \left( V_{(i)}(\sigma') \right)^N_{i=1}$. To prove the inductive step, observe first that, by the induction hypothesis, $\sigma$ and $\sigma'$ yield the same payoff under $\mu_k$ for any $k=1, \ldots, k_j-1$ (because the support of $\mu_k$ is a subset of $\Theta_{(1)}(\sigma) \cup \cdots \cup \Theta_{(j-1)}(\sigma)$).  If $j<J$, then for any $k \in \{k_{j}, \ldots, k_{j+1}-1\}$, $\sigma$ obtains expected payoff no larger than $\Pi_{(j)}$ under $\mu_k$, because the support of $\mu_k$ is a subset of $\Theta_{(1)}(\sigma) \cup \cdots \cup \Theta_{(j)}(\sigma)$. If $j=J$, then the same conclusion holds for every $k \in \{k_{J}, \ldots, K\}$, because $\Theta_{(1)}(\sigma) \cup \cdots \cup \Theta_{(J)}(\sigma)=\Theta$. Because $\sigma$ is $\mu$-optimal, it follows that, for any $\sigma'$ satisfying the induction hypothesis, either $V(\sigma, \theta)=V(\sigma',\theta)$ for all $\theta \in \Theta_{(j)}(\sigma)$ or there exists $\theta \in \Theta_{(j)}(\sigma)$ such that $V(\sigma',\theta)<\Pi_{(j)}(\sigma)=V(\sigma, \theta)$. In the latter case, $\left( V_{(i)}(\sigma) \right)^N_{i=1} >_{L} \left( V_{(i)}(\sigma') \right)^N_{i=1}$. The desired result that $\sigma$ is leximin optimal then follows by induction.

Now, suppose $|\mathcal{M}|<\infty$ and $\sigma \in \Delta(\mathcal{M})$ is leximin optimal. To show that $\sigma$ is properly robust, it suffices to exhibit a full support and strongly adversarial LPS $\mu$ against which $\sigma$ is $\mu$-optimal. Let $K=|\{V(\sigma, \theta): \theta \in \Theta\}|$ and construct $\mu=(\mu_1, \ldots, \mu_K)$ as follows. For each $k=1, \ldots, K$, let $\mu_k \in \Delta(\Theta)$ be a belief with support $\Theta_{(1)}(\sigma) \cup \cdots \cup \Theta_{(k)}(\sigma)$ such that $\sum_{\theta \in \Theta} \mu_k(\theta) V(\sigma, \theta) \geq \sum_{\theta \in \Theta} \mu_k(\theta) V(\sigma',\theta)$ for all $\sigma' \in \Delta(\mathcal{M})$. If no such belief exists, then, by Theorem 5.2.5 of \cite{blackwell1954theory} applied to the restricted decision problem with type profiles $\Theta_{(1)}(\sigma) \cup \cdots \cup \Theta_{(k)}(\sigma)$, there exists a $\sigma' \in \Delta(\mathcal{M})$ that weakly dominates $\sigma$ on $\Theta_{(1)}(\sigma) \cup \cdots \cup \Theta_{(k)}(\sigma)$.  That is, $V(\sigma',\theta) \geq V(\sigma, \theta)$ for all $\theta \in \Theta_{(1)}(\sigma) \cup \cdots \cup \Theta_{(k)}(\sigma)$ with the inequality strict for some $\theta \in \Theta_{(1)}(\sigma) \cup \cdots \cup \Theta_{(k)}(\sigma)$. But, then, there exists an $\varepsilon>0$ such that $\sigma'':= (1-\varepsilon) \circ \sigma + \varepsilon \circ \sigma'$ is strictly leximin preferred to  $\sigma$, contradicting the leximin optimality of $\sigma$. To see why such an $\varepsilon>0$ exists, observe that $V_{(i)}(\sigma)> \Pi_{(k)}(\sigma)$ for all $i > \sum^k_{j=1}|\Theta_{(j)}(\sigma)|$. So, for $\varepsilon>0$ sufficiently small, $V_{(i)}(\sigma'')> \Pi_{(k)}(\sigma)$ for all $i > \sum^k_{j=1}|\Theta_{(j)}(\sigma)|$. Moreover, for all $i \leq \sum^k_{j=1}|\Theta_{(j)}(\sigma)|$, $V_{(i)}(\sigma'') \geq V_{(i)}(\sigma)$ with the inequality strict for some $i \leq \sum^k_{j=1}|\Theta_{(j)}(\sigma)|$. Thus, $(V_{(i)}(\sigma''))^N_{i=1} >_L (V_{(i)}(\sigma))^N_{i=1}$, i.e., $\sigma$ is not leximin preferred to $\sigma''$. By construction, $\mu$ has full support and is strongly adversarial with respect to $\sigma$. Moreover, $\sigma$ is $\mu$-optimal. It follows that $\sigma$ is properly robust.

\subsection{Proof of Proposition \ref{thm_robust}}\label{proof_thm_robust}
To prove sufficiency, suppose $(Q,P) \in \mathcal{M}$ is a mechanism for which $v((Q,P),\theta) \geq v((Q,P), \theta_1)$ for all $\theta \in \Theta$. Then, the LPS $\mu=(\delta_1)$ is adversarial with respect to $(Q,P)$. If, in addition, $(Q,P)$ is efficient for the lowest type, $\theta_1$, and $P(\theta_1)=u(q^*_1, \theta_1)$, then it maximizes the seller's first-order payoff. Hence, $(Q,P)$ is $\mu$-optimal and, thus, robustly optimal.

To prove necessity, suppose that under $(Q,P) \in \mathcal{M}$ there exists a type $\theta>\theta_1$ such that $v((Q,P),\theta)<v((Q,P),\theta_1)$. Then, under any LPS $\mu$ that is adversarial with respect to $(Q,P)$, the seller's first-order payoff is strictly smaller than $v((Q,P),\theta_1)$. So, $(Q,P)$ cannot be $\mu$-optimal; the seller obtains a first-order payoff of $v((Q,P),\theta_1)$ from the mechanism $(Q',P') \in \mathcal{M}$ under which $Q'(\theta)=Q(\theta_1)$ and $P'(\theta)=P(\theta_1)$ for all $\theta \in \Theta$. 

It remains to establish efficiency for the lowest type, $\theta_1$, and $P(\theta_1)=u(q^*_1, \theta_1)$. Suppose $(Q,P) \in \mathcal{M}$ violates one of the two conditions and $v((Q,P),\theta) \geq v((Q,P),\theta_1)$ for all $\theta \in \Theta$. Then, under any LPS $\mu$ that is adversarial with respect to $(Q,P)$, the seller's first-order payoff is no larger than $v((Q,P),\theta_1)$. But, if $(Q',P')$ sets $Q'(\theta)$ equal to the Dirac measure on $q^*_1$ and $P'(\theta)=u(q^*_1, \theta_1)$ for all $\theta \in \Theta$, then $(Q',P') \in \mathcal{M}$ and $v((Q',P'),\theta) > v((Q,P),\theta_1)$ for all $\theta \in \Theta$. Hence, $(Q',P')$ yields a strictly higher first-order payoff than $(Q,P)$ against $\mu$. It follows that $(Q,P)$ is not lexicographically preferred to $(Q',P')$ and, thus, cannot be robustly optimal. 

\subsection{Proof of Proposition \ref{thm_perfect}}\label{proof_thm_perfect}

Suppose $N=2$. Because every perfectly robust mechanism is robust, Proposition \ref{thm_robust} establishes that any perfectly robust mechanism must maximize profit from type $\theta_1$. Among all mechanisms with this property, the efficient and maximal mechanism maximizes $v((Q,P),\theta_2)$. Hence, any other mechanism is weakly dominated and therefore, by Proposition \ref{perfect_admissibility}, cannot be perfectly robust. To establish that the efficient and maximal mechanism is, in fact, perfectly robust, it suffices to observe that it is $\mu$-optimal with respect to the adversarial and full support LPS $\mu=(\delta_1, \delta_2)$.

It is next shown that, if $(Q,P) \in \mathcal{M}$ is perfectly robust and deterministic, then $P$ satisfies \eqref{rev}. By construction, if $P$ satisfies \eqref{rev}, then for any transfer rule $P' \neq P$ that implements $Q$, $P(\theta) \geq P'(\theta)$ for all $\theta \in \Theta$ with the inequality strict for some $\theta' \in \Theta$. It follows that $v((Q,P),\theta) \geq v((Q,P'),\theta)$ with the inequality strict for some $\theta \in \Theta$. That is, $(Q,P)$ weakly dominates $(Q,P')$. Because admissibility is a necessary condition for perfect robustness (Proposition \ref{perfect_admissibility}), it follows that $P$ must satisfy \eqref{rev} for $(Q,P)$ to be perfectly robust.

It remains to show that, if $(Q,P) \in \mathcal{M}$ is perfectly robust and deterministic, then $Q$ is efficient for types $\theta_1$ and $\theta_N$. Because every perfectly robust mechanism is robust, Proposition \ref{thm_robust} ensures that $Q$ is efficient for type $\theta_1$. Suppose, towards contradiction, that $Q$ is not efficient for type $\theta_N$. Define an allocation rule $Q'$ by setting $Q'(\theta_N)$ equal to the Dirac measure on $q^*_N$ and, for each $\theta \neq \theta_N$, setting $Q'(\theta)= Q(\theta)$ if $Q(\theta)$ is a Dirac measure on a quality strictly below $q^*_N$, and setting $Q'(\theta)$ equal to the Dirac measure on $q^*_N$ otherwise. Let $P'$ satisfy \eqref{rev} with $Q'$ replacing $Q$. Then $(Q',P') \in \mathcal{M}$ weakly dominates $(Q,P)$: profit from $\theta_N$ strictly increases, while profit from all other types remains at least the same. Hence, by Proposition \ref{perfect_admissibility}, $(Q,P)$ cannot be perfectly robust.

\subsection{Proof of Proposition \ref{thm_proprobust}}\label{proof_thm_proprobust}

Suppose $(Q,P) \in \mathcal{M}$ is the efficient and maximal mechanism. To prove that it is properly robust, consider the LPS $\mu=(\delta_1, \delta_2, \ldots, \delta_N)$. By construction, $\mu$ has full support.  $\mu$ is also strongly adversarial with respect to $(Q,P)$. To see why, take any $\theta, \theta' \in \Theta$. By construction of $\mu$, $\theta \geq_\mu \theta'$ implies $\theta' \geq \theta$. Because the seller's payoff under $(Q,P)$ is increasing in the buyer's type, $\theta' \geq \theta$ implies $v((Q,P), \theta) \leq v((Q,P), \theta')$. Hence,  $\theta \geq_\mu \theta'$ implies $v((Q,P), \theta) \leq v((Q,P), \theta')$, so $\mu$ is strongly adversarial with respect to $(Q,P)$. To show that $(Q,P)$ is $\mu$-optimal (and, in fact, uniquely optimal against $\mu$), proceed by induction on the vector of payoffs obtained from an arbitrary mechanism $(Q',P') \in \mathcal{M}$. For the base case, individual rationality and the uniqueness of $q^*_1$ in maximizing surplus from $\theta_1$ imply that $(Q(\theta_1), P(\theta_1))$ uniquely maximizes profit against $\delta_1$. Hence, $(Q',P')$ cannot obtain a higher first-order payoff than $(Q,P)$ against $\mu$ and equality requires $(Q'(\theta_1), P'(\theta_1))=(Q(\theta_1), P(\theta_1))$. Now, suppose $Q'$ is efficient for types $\theta_1, \ldots, \theta_k$ and $P'$ is revenue maximizing given $Q'$ for types $\theta_1, \ldots, \theta_k$. That is,
\begin{equation}
\begin{aligned}
P'(\theta_1) = u(q^*_1, \theta_1) \quad \text{and} \quad 
P'(\theta_j) = u(q^*_1, \theta_1)+ \sum^j_{i=2} \left[ u(q^*_i, \theta_i)- u(q^*_{i-1}, \theta_{i})\right] \quad \text{for $j \in \{2,...,k\}$.}
\end{aligned}\notag
\end{equation}
Any mechanism in this class that maximizes the seller's $(k+1)$-th order payoff cannot yield a higher payoff than the solution to the relaxed problem
\begin{equation}
\begin{aligned}
&\max_{Q \in \Delta(\mathcal{Q}),p \in \mathbb{R}} \quad p-C(Q) \\
&\text{subject to}\\
& U(Q, \theta_{k+1})-p \geq \bar{u} := u(q^*_k, \theta_{k+1})- P(\theta_k).
\end{aligned} \label{induction}
\end{equation}
In any solution to \eqref{induction}, $p$ is determined by the binding constraint. Substituting the binding constraint into the objective function and eliminating the constant $\bar{u}$ yields the surplus maximization problem
\begin{equation}
\begin{aligned}
&\max_{Q \in \Delta(\mathcal{Q})} \quad U(Q, \theta_{k+1})-C(Q),
\end{aligned} \notag
\end{equation}
whose unique solution is the Dirac measure on $q^*_{k+1}$.
The binding constraint in \eqref{induction} then uniquely
determines $p=P(\theta_{k+1})$. It follows that, among mechanisms
coinciding with $(Q,P)$ at types $\theta_1,\ldots,\theta_k$,
any mechanism maximizing the seller's $(k+1)$-th order payoff
must also coincide with $(Q,P)$ at $\theta_{k+1}$.
The unique $\mu$-optimality of $(Q,P)$ follows from induction.
Because $\mu$ has full support and is strongly adversarial
with respect to $(Q,P)$, this mechanism is properly robust.

To establish uniqueness, let $(\hat Q,\hat P)\in\mathcal M$
be any properly robust mechanism. It is shown that
\begin{equation}
v((\hat Q,\hat P),\theta)=v((Q,P),\theta)
\quad\text{for all }\theta\in\Theta, \label{lex_equal} \end{equation} where $(Q,P)$ is the efficient and maximal mechanism. If true, then $(\hat Q,\hat P)$ is optimal against
$\mu=(\delta_1, \delta_2, \ldots, \delta_N)$. The unique $\mu$-optimality of $(Q,P)$ established in the previous paragraph then implies
$(\hat Q,\hat P)=(Q,P)$, thereby establishing the result. To establish \eqref{lex_equal}, note that, by Proposition
\ref{proper_leximin}, both $(Q,P)$ and $(\hat Q,\hat P)$ are
leximin optimal and therefore have the same ordered payoff
vector. Suppose, towards contradiction, that their payoffs differ at some type and let
\[
\alpha:=\min_{\theta:\,v((Q,P),\theta)\neq v((\hat Q,\hat P),\theta)}
\min\{v((Q,P),\theta),v((\hat Q,\hat P),\theta)\}.
\]
Because incentive compatibility and individual rationality
are preserved by averaging allocation lotteries and transfers,
there exists a mechanism $(\bar Q,\bar P)\in\mathcal M$ that
is payoff equivalent to the lottery
$\frac12\circ(Q,P)+\frac12\circ(\hat Q,\hat P)$ and satisfies
\[
v((\bar Q,\bar P),\theta)
=\tfrac12 v((Q,P),\theta)+\tfrac12 v((\hat Q,\hat P),\theta)
\quad\text{for all }\theta\in\Theta.
\]
At every type where the original payoffs differ, both payoffs
are at least $\alpha$ and their average is strictly greater than $\alpha$.
At every other type, the payoff is unchanged. Since the original
mechanisms have the same ordered payoff vector, their average
therefore leaves all payoffs below $\alpha$ unchanged and strictly
reduces the number of payoffs equal to $\alpha$. The ordered payoff
vector associated with $(\bar Q,\bar P)$ is thus strictly higher in the lexicographic order, contradicting the supposed leximin optimality of $(Q,P)$ and $(\hat Q, \hat P)$.

\subsection{Proof of Proposition \ref{thm_auction_robust}}\label{proof_thm_auction_robust}

To prove sufficiency, suppose $(Q,P) \in \mathcal{M}$ is a mechanism for which $v((Q,P),\theta) \geq v((Q,P), \underline{\theta})$ for all $\theta \in \Theta$. Then, the LPS $\mu=(\delta_{\underline{\theta}})$ is adversarial with respect to $(Q,P)$. If, in addition, $(Q,P)$ is efficient for the lowest type profile and extracts full surplus under that profile, then it maximizes the auctioneer's first-order payoff. Hence, $(Q,P)$ is $\mu$-optimal and, thus, robustly optimal.

To prove necessity, suppose that under $(Q,P) \in \mathcal{M}$ there exists a type profile $\theta \neq \underline{\theta}$ such that $v((Q,P),\theta)<v((Q,P),\underline{\theta})$. Then, under any LPS $\mu$ that is adversarial with respect to $(Q,P)$, the auctioneer's first-order payoff is strictly smaller than $v((Q,P),\underline{\theta})$. So, $(Q,P)$ cannot be $\mu$-optimal; the auctioneer obtains a first-order payoff of $v((Q,P),\underline{\theta})$ from the mechanism $(\hat{Q},\hat{P}) \in \mathcal{M}$ under which $\hat{Q}^i(\theta)=Q^i(\underline{\theta})$ and $\hat{P}^i(\theta)=P^i(\underline{\theta})$ for all $i=1, \ldots, I$ and $\theta \in \Theta$. 

It remains to establish the necessity of efficiency and full surplus extraction for the lowest type profile $\underline{\theta}$. Suppose $(Q,P) \in \mathcal{M}$ violates any of these two conditions and $v((Q,P),\theta) \geq v((Q,P),\underline{\theta})$ for all $\theta \in \Theta$. Then, under any LPS $\mu$ that is adversarial with respect to $(Q,P)$, the seller's first-order payoff is no larger than $v((Q,P),\underline{\theta})$. But, if $(\hat{Q},\hat{P}) \in \mathcal{M}$ always allocates the good to, say, agent $1$ with probability one at a price equal to $\theta_1$, then $v((\hat{Q},\hat{P}),\theta) > v((Q,P),\underline{\theta})$ for all $\theta \in \Theta$. (Formally, for all $\theta \in \Theta$, set $\hat{Q}^{1}(\theta)=1$ and $\hat{Q}^{j}(\theta)=0$ for all $j \neq 1$. Then, set $\hat{P}^{1}(\theta)=\theta_1$ and $\hat{P}^{j}(\theta)=0$ for all $j \neq 1$.) Hence, $(\hat{Q},\hat{P})$ yields a strictly higher first-order payoff than $(Q,P)$ against $\mu$. It follows that $(Q,P)$ is not lexicographically preferred to $(\hat{Q},\hat{P})$ and, thus, cannot be robustly optimal.

\subsection{Proof of Proposition \ref{thm_auction_perfect}}\label{proof_thm_auction_perfect}

It is first shown that, if $(Q,P) \in \mathcal{M}$ is perfectly robust, then $P$ satisfies \eqref{rev_auction}. By construction, if $P$ satisfies \eqref{rev_auction}, then for any transfer rule $P' \neq P$ that implements $Q$, $P(\theta) \geq P'(\theta)$ for all $\theta \in \Theta$ with the inequality strict for some $\theta' \in \Theta$. It follows that $v((Q,P),\theta) \geq v((Q,P'),\theta)$ with the inequality strict for some $\theta \in \Theta$. That is, $(Q,P)$ weakly dominates $(Q,P')$. Because admissibility is a necessary condition for perfect robustness (Proposition \ref{perfect_admissibility}), it follows that $P$ must satisfy \eqref{rev_auction} for $(Q,P)$ to be perfectly robust.

Because every perfectly robust mechanism is robust, Proposition \ref{thm_auction_robust} already ensures that any perfectly robust mechanism $(Q,P) \in \mathcal{M}$ is efficient for type profile $\underline{\theta}$ (and, therefore, for any type profile $\theta \in \Theta$ satisfying $\max(\theta)=\theta_1$). To show that any perfectly robust mechanism $(Q,P) \in \mathcal{M}$ is efficient for any type profile $\theta \in \Theta$ satisfying $\max(\theta)=\theta_N$, take any perfectly robust mechanism $(Q,P) \in \mathcal{M}$. It has already been shown that $P$ must satisfy \eqref{rev_auction}. Suppose, towards contradiction, that there exists a type profile $\theta \in \Theta$ such that $\max(\theta)=\theta_N$ and either $\sum_{i=1}^I Q^i(\theta)<1$ or $Q^i(\theta)>0$ for some $i$ satisfying $\theta^i<\theta_N$. Let $j$ be an agent for whom $\theta^j=\theta_N$. Define a mechanism $(\hat Q,\hat P)\in\mathcal M$ as follows. For every type profile $\tilde\theta$ satisfying $\tilde\theta^j=\theta_N$, set $\hat Q^j(\tilde\theta)=1$ and
$\hat Q^i(\tilde\theta)=0 \ \text{for all } i\neq j$.
For every type profile $\tilde\theta$ for which $\tilde\theta^j \neq \theta_N$, set $\hat Q(\tilde\theta)=Q(\tilde\theta)$. Finally, let $\hat P$ satisfy \eqref{rev_auction} with $\hat Q$ replacing $Q$. Note that $\hat Q$ remains increasing in each agent's type. Hence, $(\hat Q,\hat P)\in\mathcal M$. Moreover, revenue may only change at a type profile $\tilde\theta$ with $\tilde\theta^j=\theta_N$. But, \eqref{rev_auction} implies
\[
v((\hat Q,\hat P),\tilde\theta)-v((Q,P),\tilde\theta)
\geq
\sum_{i=1}^I \tilde\theta^i\bigl(\hat Q^i(\tilde\theta)-Q^i(\tilde\theta)\bigr).
\]
Therefore,
\[
v((\hat Q,\hat P),\tilde\theta)-v((Q,P),\tilde\theta)
\geq
\theta_N\Bigl(1-\sum_{i=1}^I Q^i(\tilde\theta)\Bigr)
+
\sum_{i\neq j}(\theta_N-\tilde\theta^i)Q^i(\tilde\theta)\ge 0.
\]
At the original type profile $\theta$, the inequality is strict. Hence $(\hat Q,\hat P)$ weakly dominates $(Q,P)$, contradicting Proposition \ref{perfect_admissibility}. It follows that any perfectly robust mechanism must be efficient at every type profile $\theta$ satisfying $\max(\theta)=\theta_N$.

Suppose $N=2$. It has been established that, if $(Q,P) \in \mathcal{M}$ is perfectly robust, then $Q$ is efficient for every type profile $\theta \in \Theta$ satisfying either $\max(\theta)=\theta_1$ or $\max(\theta)=\theta_N$. If $N=2$, then every type profile $\theta \in \Theta$ satisfies $\max(\theta)=\theta_1$ or $\max(\theta)=\theta_N$. Hence, $Q$ must be efficient. Having already established that $P$ must satisfy \eqref{rev_auction}, it follows that every perfectly robust mechanism is efficient and maximal. To establish that any efficient and maximal mechanism $(Q,P) \in \mathcal{M}$ is perfectly robust, consider the adversarial and full support LPS $\mu=(\delta_{\underline{\theta}}, \mu_2, \delta_{\overline{\theta}})$, where $\mu_2$ is the uniform probability measure over the set of type profiles $\theta \in \Theta \backslash \{\underline{\theta},\overline{\theta}\}$. It is straightforward to verify that $(Q,P)$ is $\mu$-optimal.

\subsection{Proof of Proposition \ref{thm_auction_proper}}\label{proof_thm_auction_proper}

\subsubsection{Construction of a full support and strongly adversarial LPS}

Fix an efficient and maximal mechanism $(Q,P) \in \mathcal{M}$ that breaks ties uniformly for all type profiles $\theta$ with $\max(\theta)<\theta_N$. Construct a full support and strongly adversarial LPS $\mu$ of length $1+(N-1)(2(I-1)+1)$ as follows. Let $\mu_1$ be the point mass on $\underline{\theta}$. For each $k=2, \ldots, N$ and $\ell=1, \ldots, I-1$, let $b(k)=2+(k-2)(2(I-1)+1)$; let $\Theta^k_\ell$ be the set of type profiles $\theta$ satisfying $\max(\theta)=\theta_k$, $|\underset{i \in \mathcal{I}}{\argmax}~ \theta^i|=1$, and $|\{i \in \mathcal{I}: \theta^i= \theta_{k-1}\}|=\ell$; and let $\hat{\Theta}^k_\ell$ be the set of type profiles $\theta$ satisfying $\max(\theta) > \theta_k$, $|\underset{i \in \mathcal{I}}{\argmax}~ \theta^i|=1$, $\theta_{(I-1)}=\theta_{k-1}$ (where $\theta_{(I-1)}$ is the second-highest value in $\theta$), and $|\{i \in \mathcal{I}: \theta^i= \theta_{k-1}\}|=\ell$. If $\Theta^k_\ell$ is non-empty, define $\mu_{b(k)+2(\ell-1)}$ to be the uniform probability measure with support $\Theta^k_\ell$. Otherwise, set $\mu_{b(k)+2(\ell-1)}=\mu_{b(k)+2(\ell-1)-1}$. If $\hat{\Theta}^k_\ell$ is non-empty, define $\mu_{b(k)+2(\ell-1)+1}$ to be the uniform probability measure with support $\hat{\Theta}^k_\ell$. Otherwise, set $\mu_{b(k)+2(\ell-1)+1}=\mu_{b(k)+2(\ell-1)}$. Finally, let $\mu_{b(k)+2(I-1)}$ be the uniform probability measure over the set of type profiles with $\max(\theta)=\theta_k$ and $|\underset{i \in \mathcal{I}}{\argmax}~ \theta^i| > 1$.

It is immediate that $\mu$ has full support. Towards showing that $\mu$ is strongly adversarial with respect to $(Q,P)$, define
\[
v_{k,\ell}:=\left(\frac{1}{\ell+1}\right)\theta_{k-1}
+\left(1-\frac{1}{\ell+1}\right)\theta_k.
\]
Let $(v_{(k)})^{N^I}_{k=1}$ denote the $N^I$-dimensional vector of payoff order statistics associated with any efficient and maximal mechanism $(Q,P) \in \mathcal{M}$ that breaks ties uniformly for all type profiles $\theta$ with $\max(\theta)<\theta_N$. When $N=2$, this vector takes on three values
\[
\theta_1<v_{2,I-1}<\theta_2.
\]
For $N>2$, the vector takes on values
\[
\theta_1<v_{2,I-1}<\theta_2<v_{3,1}<\cdots<v_{3,I-1}
<\cdots<\theta_{N-1}<v_{N,1}<\cdots<v_{N,I-1}<\theta_N.
\]
To see that $\mu$ is strongly adversarial, observe that each belief from $\mu_1$ through $\mu_{b(2)+2(I-2)-1}$ is the point mass on $\underline{\theta}$, at which $(Q,P)$ yields revenue $\theta_1$. For each $(k,\ell)$ such that $k=2$ and $\ell=I-1$, or $k>2$ and $\ell=1,\ldots,I-1$, $(Q,P)$ yields revenue $v_{k,\ell}$ at every type profile in the support of either $\mu_{b(k)+2(\ell-1)}$ or $\mu_{b(k)+2(\ell-1)+1}$. Finally, it yields revenue $\theta_k$ at every type profile in the support of $\mu_{b(k)+2(I-1)}$. The displayed ordering of revenue levels therefore implies that $\mu$ is strongly adversarial with respect to $(Q,P)$.

\subsubsection{$\mu$-optimality of candidate mechanism}\label{proof_thm_auction_proper_part2}

To show that $(Q,P)$ is $\mu$-optimal and therefore properly robust, consider any mechanism $(\hat Q,\hat P) \in \mathcal{M}$ in which $\hat P$ satisfies \eqref{rev_auction} with $\hat Q$ replacing $Q$ (it is without loss of generality to fix such a transfer rule because it maximizes revenue pointwise). The point mass on $\underline{\theta}$ is repeated through $\mu_{b(2)+2(I-2)-1}$. Hence, in order to be $\mu$-optimal, $(\hat Q,\hat P)$ must maximize profit under $\underline{\theta}$. For each $\theta \in \Theta^k_\ell$, let $i^*_\theta$ denote the agent $i$ with $\theta^i=\theta_k$. For $(k,\ell)$ such that $k=2$ and $\ell=I-1$, or $k>2$ and $\ell=1,\ldots,I-1$, any mechanism $(\hat Q,\hat P) \in \mathcal{M}$ with
$\sum_{j\in\underset{i\in \mathcal{I}}{\argmax}\,\theta^i}\hat Q^j(\theta)=1$
for any type profile $\theta$ satisfying $\max(\theta)=\theta_{k-1}$ obtains no higher revenue than $(Q,P)$ against $\mu_{b(k)+2(\ell-1)}$:
\[
\begin{aligned}
\sum_{\theta\in\Theta^k_\ell}
\frac{1}{|\Theta^k_\ell|}
\left[\sum_{i=1}^I \hat P^i(\theta)\right]
&\leq
\sum_{\theta\in\Theta^k_\ell}
\frac{
	\hat Q^{i^*_\theta}(\theta_{k-1},\theta^{-i^*_\theta})\theta_{k-1}
	+
	\left(1-\hat Q^{i^*_\theta}(\theta_{k-1},\theta^{-i^*_\theta})\right)\theta_k
}{|\Theta^k_\ell|}\\
&=
\theta_k-(\theta_k-\theta_{k-1})
\frac{
	\sum_{\theta\in\Theta^k_\ell}
	\hat Q^{i^*_\theta}(\theta_{k-1},\theta^{-i^*_\theta})
}{|\Theta^k_\ell|}\\
&=
\theta_k-(\theta_k-\theta_{k-1})
\frac{\frac{1}{\ell+1}|\Theta^k_\ell|}{|\Theta^k_\ell|}\\
&=
\left(\frac{1}{\ell+1}\right)\theta_{k-1}
+
\left(1-\frac{1}{\ell+1}\right)\theta_k.
\end{aligned}
\]
The first inequality holds with equality if and only if, for any type profile $\theta\in\Theta^k_\ell$, $\hat Q^{i^*_\theta}(\theta)=1$. The second equality holds because
$\sum_{j\in\underset{i\in \mathcal{I}}{\argmax}\,\theta^i}\hat Q^j(\theta)=1$
for any type profile $\theta$ satisfying $\max(\theta)=\theta_{k-1}$. For each $\theta\in\Theta^k_\ell$, lowering the unique $\theta_k$ agent to $\theta_{k-1}$ produces a profile in which there are exactly $\ell+1$ agents with type $\theta_{k-1}$, at which the maintained condition ensures that allocation probabilities across these $\ell+1$ agents sum to one. Each such lowered profile is generated exactly $\ell+1$ times---once for each possible choice of the agent designated as $i^*_\theta$. Consequently, summing
$\hat Q^{i^*_\theta}(\theta_{k-1},\theta^{-i^*_\theta})$
over $\theta\in\Theta^k_\ell$ yields total mass
$\frac{1}{\ell+1}|\Theta^k_\ell|$. Now, consider $\mu_{b(k)+2(\ell-1)+1}$. Suppose $(\hat Q,\hat P)$ satisfies the condition that, for any type profile $\theta\in\Theta^k_\ell$, $\hat Q^{i^*_\theta}(\theta)=1$. Then, it follows from monotonicity, $\hat P$ satisfying \eqref{rev_auction} with $\hat Q$ replacing $Q$, and the same averaging argument that the expected revenue from $(\hat Q,\hat P)$ against $\mu_{b(k)+2(\ell-1)+1}$ is exactly $v_{k,\ell}$. Finally, for $k=2,\ldots,N$, observe that no mechanism in $\mathcal{M}$ can obtain a higher payoff than $\theta_k$ against $\mu_{b(k)+2(I-1)}$ because $\theta_k$ is the highest type in all type profiles in the support of $\mu_{b(k)+2(I-1)}$. And, any $(\hat Q,\hat P)$ that obtains $\theta_k$ must set
$\sum_{i\in\arg\max_j\theta^j}\hat Q^i(\theta)=1$
for any $\theta\in\operatorname{supp}(\mu_{b(k)+2(I-1)})$. It follows from iterative optimization that $(Q,P)$ must be $\mu$-optimal. Moreover, the equality conditions above and monotonicity imply that every $\mu$-optimal mechanism is efficient; maximality follows from the full support of $\mu$.

\subsubsection{Converse}

By Proposition \ref{proper_leximin}, any properly robust mechanism
$(\hat Q,\hat P) \in \mathcal{M}$ must have a corresponding vector of payoff
order statistics $(\hat v_{(k)})^{N^I}_{k=1}$ such that
\[
(\hat v_{(k)})^{N^I}_{k=1}
\geq_L
(v_{(k)})^{N^I}_{k=1},
\]
where $(v_{(k)})^{N^I}_{k=1}$ is the vector of payoff order statistics
associated with the efficient and maximal mechanism $(Q,P)$ that breaks
ties uniformly for all type profiles $\theta$ with
$\max(\theta)<\theta_N$. Since $(Q,P)$ is also properly robust, both
mechanisms are leximin optimal, and the displayed inequality must hold
with equality. Moreover, their payoffs must coincide at every type profile:
\[
v((\hat Q,\hat P),\theta)=v((Q,P),\theta)
\qquad\text{for all $\theta\in\Theta$.}
\]
To see this, suppose otherwise and let $\alpha$ be the smallest payoff, under
either mechanism, at a profile where their payoffs differ. Because
$\mathcal{M}$ is convex, the average mechanism putting half weight on $(\hat Q, \hat P)$ and half weight on $(Q,P)$ is feasible.  Its payoffs below $\alpha$ are unchanged. Since the two original mechanisms have the same ordered
payoff vector, their average has strictly fewer payoffs equal to $\alpha$.
It therefore has a strictly higher ordered payoff vector, contradicting
leximin optimality. Because $(Q,P)$ is $\mu$-optimal, this pointwise payoff equality implies
that $(\hat Q,\hat P)$ is optimal against the same $\mu$. The conclusion
of Section \ref{proof_thm_auction_proper_part2} therefore implies that $(\hat Q,\hat P)$ is efficient and maximal.

Next, fix $k\in\{1,\ldots,N-1\}$ and $\ell\in\{1,\ldots,I-1\}$, and
consider a profile $\theta$ with $\max(\theta)=\theta_k$ and exactly
$I-\ell+1$ agents of type $\theta_k$. Uniform tie-breaking at such a
profile requires that each of these $I-\ell+1$ agents receive the good
with probability $1/(I-\ell+1)$. Suppose, towards contradiction, that
ties are not broken uniformly at $\theta$. Since $\hat Q$ is efficient,
there exists one of the agents with type $\theta_k$, say agent $i$, for
whom
\[
\hat Q^i(\theta)>\frac{1}{I-\ell+1}.
\]
Let $\hat\theta$ be the profile obtained from $\theta$ by replacing agent
$i$'s type $\theta_k$ with $\theta_{k+1}$. Because $\hat P$ satisfies
\eqref{rev_auction}, the auctioneer's revenue at $\hat\theta$ is no higher
than
\[
\hat Q^i(\theta)\theta_k
+\left(1-\hat Q^i(\theta)\right)\theta_{k+1}
<
v_{k+1,I-\ell}
=
v((Q,P),\hat\theta).
\]
This contradicts the pointwise payoff equality established above.
Therefore, ties must be broken uniformly at every profile whose highest
type is below $\theta_N$. In conclusion, any properly robust mechanism
$(\hat Q,\hat P)\in\mathcal{M}$ must be efficient and maximal, with ties
broken uniformly for all type profiles $\theta$ satisfying
$\max(\theta)<\theta_N$.

\subsection{Proof of Proposition \ref{thm_pub_robust}}\label{proof_thm_pub_robust}

To prove sufficiency, suppose $(Q,P) \in \mathcal{M}$ is a mechanism such that,  for all $\theta \in \Theta$, $\left(\sum^I_{i=1} P^i(\theta) \right)-Q(\theta) c \geq 0$. Then, the LPS $\mu=(\delta_{\underline{\theta}})$ is adversarial with respect to $(Q,P)$ and the mechanism is $\mu$-optimal. To prove necessity, suppose towards contradiction that there exists $\theta \in \Theta$ such that $\left(\sum^I_{i=1} P^i(\theta) \right)-Q(\theta) c < 0$.
Then, under any LPS $\mu$ that is adversarial with respect to $(Q,P)$, the firm obtains a negative first-order payoff. But then, $(Q,P)$ cannot be $\mu$-optimal; the firm obtains a first-order payoff of zero under the mechanism in which, for all $\theta \in \Theta$ and all $i$, $Q(\theta)=P^i(\theta)=0$.

\subsection{Proof of Proposition \ref{thm_pub_perfect}}\label{proof_thm_pub_perfect}

It is first shown that, if $(Q,P) \in \mathcal{M}$ is perfectly robust, then $P$ satisfies \eqref{rev_pub}. By construction, if $P$ satisfies \eqref{rev_pub}, then for any transfer rule $P' \neq P$ that implements $Q$, $P(\theta) \geq P'(\theta)$ for all $\theta \in \Theta$ with the inequality strict for some $\theta' \in \Theta$. It follows that $v((Q,P),\theta) \geq v((Q,P'),\theta)$ with the inequality strict for some $\theta \in \Theta$. That is, $(Q,P)$ weakly dominates $(Q,P')$. Because admissibility is a necessary condition for perfect robustness (Proposition \ref{perfect_admissibility}), it follows that $P$ must satisfy \eqref{rev_pub} for $(Q,P)$ to be perfectly robust.

It is next shown that the allocation rule $Q$ is efficient for any type profile $\theta \in \Theta$ such that $\sum^I_{i=1} \theta^i \leq c$ or $\theta=\overline{\theta}$. Because every perfectly robust mechanism is robust, Proposition \ref{thm_pub_robust} already ensures that any perfectly robust mechanism $(Q,P) \in \mathcal{M}$ is efficient for any type profile $\theta \in \Theta$ such that $\sum^I_{i=1} \theta^i < c$. In particular, if $Q(\theta)>0$ for such a type profile, then, by the individual rationality constraints, the firm obtains a payoff no higher than the negative number $Q(\theta) \left(\sum^I_{i=1} \theta^i-c \right)<0$. Moreover, for any mechanism $(Q,P) \in \mathcal{M}$ with $P$ satisfying \eqref{rev_pub} such that $Q(\theta)>0$ for some type profile $\theta \in \Theta$ satisfying $\sum^I_{i=1} \theta^i =c$ and $Q(\hat \theta)=0$ for any type profile $\hat \theta \in \Theta$ such that $\sum^I_{i=1} \hat \theta^i < c$, there exists another mechanism $(\hat{Q}, \hat{P}) \in \mathcal{M}$ that weakly dominates $(Q,P)$. In particular, set $\hat{Q}(\theta)=0$ and $\hat{Q}(\hat{\theta})=Q(\hat{\theta})$ for $\hat{\theta} \neq \theta$ and define $\hat{P}$ using \eqref{rev_pub} replacing $Q$ with $\hat{Q}$. At $\theta$, every profile obtained by lowering a high-type agent's report has strictly negative surplus and hence zero provision. Thus, by \eqref{rev_pub},
\[
v((Q,P),\theta)
=Q(\theta)\left(\sum_{j=1}^I\theta^j-c\right)
=0=v((\hat Q,\hat P),\theta).
\]
Because $c<I\theta_h$, there exists an agent $i$ for whom $\theta^i=\theta_\ell$. Under the type profile $\tilde{\theta}$ in which $\tilde{\theta}^j=\theta^j$ for all $j \neq i$ and $\tilde{\theta}^i=\theta_h$, the firm's payoff strictly increases. For every other type profile, the firm's payoff does not strictly decrease. It follows that $(\hat{Q},\hat{P})$ weakly dominates $(Q,P)$. Hence, $(Q,P)$ could not have been perfectly robust by Proposition \ref{perfect_admissibility}. Finally, if $(Q,P) \in \mathcal{M}$ satisfied $Q(\overline{\theta})<1$, then the mechanism $(\hat{Q},\hat{P}) \in \mathcal{M}$ defined by $\hat{Q}(\overline{\theta})=1$ and $\hat{Q}(\theta)=Q(\theta)$ for all $\theta \neq \overline{\theta}$, with $\hat{P}$ satisfying \eqref{rev_pub} replacing $Q$ with $\hat{Q}$, weakly dominates $(Q,P)$. Hence, $(Q,P)$ could not have been perfectly robust by Proposition \ref{perfect_admissibility}.

To see that the efficient and maximal mechanism is perfectly robust when $c \in [(I-1)\theta_h+\theta_\ell,I\theta_h)$, observe that it is optimal against any adversarial and full support LPS $\mu=(\mu_1,\delta_{\overline{\theta}})$, where $\mu_1$ is a full support distribution over $\Theta \backslash \{\overline{\theta}\}$. It is the unique perfectly robust mechanism by the necessary conditions for perfection previously established.

\subsection{Proof of Proposition \ref{thm_pub_proper}}\label{proof_thm_pub_proper}

The proof proceeds by first showing that any properly robust mechanism must be anonymous,  i.e., $Q(\theta)=Q(\hat{\theta})$ if $|\{i: \theta^i=\theta_h\}|=|\{i: \hat{\theta}^i=\theta_h\}|$. Then, the conditions provided for $(Q,P) \in \mathcal{M}$ in the statement of the proposition are shown to be both necessary and sufficient for proper robustness.

\subsubsection{Anonymity}\label{pub_anon}

Let $(Q,P)\in\mathcal M$ be properly robust. By Proposition
\ref{proper_leximin}, $(Q,P)$ is leximin optimal. This observation
is used to show that $v((Q,P),\theta)=v((Q,P),\hat\theta)$ whenever
$|\{i:\theta^i=\theta_h\}|=|\{i:\hat\theta^i=\theta_h\}|$.
Suppose towards contradiction that there exist such profiles
$\theta,\hat\theta$ with different payoffs. Pair each agent who has type $\theta_h$ in $\theta$ and type
$\theta_\ell$ in $\hat\theta$ with an agent whose types are reversed.
Let $\pi$ swap the agents in each pair and leave all other agents
fixed, so that $\theta^{\pi(i)}=\hat\theta^i$ for all $i$. Define
$(Q_\pi,P_\pi)\in\mathcal M$ by
$Q_\pi(\theta')=Q(\theta'_\pi)$ and
$P_\pi^i(\theta')=P^{\pi(i)}(\theta'_\pi)$ for each type profile
$\theta'$. This mechanism merely permutes the type-profile payoffs of $(Q,P)$
and therefore has the same ordered payoff vector, so it is also
leximin optimal. If their payoffs differed at any profile, their
average would be feasible and, by the averaging argument in the
proof of Proposition \ref{thm_auction_proper}, would have a strictly
higher ordered payoff vector. This would contradict leximin
optimality. Their payoffs must therefore coincide at every type
profile. But then
\[
v((Q_\pi,P_\pi),\theta)=v((Q,P),\hat\theta)
\neq v((Q,P),\theta),
\]
a contradiction.

The preceding paragraph establishes that in any properly robust mechanism $(Q,P) \in \mathcal{M}$, for each $k \in \{0, \ldots, I\}$, there exists a number $\pi_k$ such that $v((Q,P),\theta)=\pi_k$ for any $\theta$ with $|\{i: \theta^i=\theta_h\}|=k$. It remains to show that $Q(\theta)=Q(\hat{\theta})$ if $|\{i: \theta^i=\theta_h\}|=|\{i: \hat{\theta}^i=\theta_h\}|$. Let $\theta_S$ denote the type profile in which the agents $S \subseteq \mathcal{I}$ have type $\theta_h$ and all others have type $\theta_\ell$. Because any properly robust mechanism is perfectly robust, Proposition \ref{thm_pub_perfect} implies that $P$ must be determined by \eqref{rev_pub}. So, if $|\{i : \theta^i_S=\theta_h\}|=k$, then the firm's payoff under $\theta_S$ is
\begin{equation}
    v((Q,P),\theta_S) = Q(\theta_S) S_{k}- (\theta_h-\theta_\ell) \sum_{i \in S} Q(\theta_{S \backslash \{i\}})=\pi_k. \notag
\end{equation}
Because $(Q,P)$ is perfectly robust, the argument in the proof of
Proposition \ref{thm_pub_perfect} implies that $Q(\theta)=0$
whenever $\sum_i\theta^i\leq c$. Thus, allocation anonymity holds
for $k<k^*$.  Now, consider the case in which $k=k^*$. Because $Q(\theta)=0$ for any $\theta$ such that $|\{i : \theta^i=\theta_h\}|<k$, $v((Q,P),\theta)= Q(\theta) S_{k^*}=\pi_{k^*}$ if $|\{i : \theta^i=\theta_h\}|=k^*$. Hence, $Q(\theta)=\pi_{k^*} / S_{k^*}$ for any $\theta$ satisfying $|\{i : \theta^i=\theta_h\}|=k^*$, i.e., $Q$ only depends on $\theta$ through $|\{i : \theta^i=\theta_h\}|=k^*$. For the inductive step, let $k>k^*$ and suppose $Q(\theta)=Q(\hat{\theta})=Q_{k-1}$ if $|\{i: \theta^i=\theta_h\}|=|\{i: \hat{\theta}^i=\theta_h\}|=k-1$. Let $S \subseteq \mathcal{I}$ satisfy $|S|=k$. By the induction hypothesis,
\[v((Q,P), \theta_S)= Q(\theta_S) S_k- (\theta_h-\theta_\ell) k Q_{k-1}=\pi_k. \]
Hence,
\[Q(\theta_S)= \frac{\pi_k+(\theta_h-\theta_\ell) k Q_{k-1}}{S_k}:= Q_k.\] Thus, $Q(\theta)=Q(\hat{\theta})=Q_k$ if $|\{i: \theta^i=\theta_h\}|=|\{i: \hat{\theta}^i=\theta_h\}|=k$. The desired result follows from induction.

\subsubsection{Necessity}\label{pub_neccesity}

Because any properly robust mechanism is perfectly robust, Proposition \ref{thm_pub_perfect} implies that $P$ must be determined by \eqref{rev_pub}. Section \ref{pub_anon} established that, in any properly robust mechanism $(Q,P) \in \mathcal{M}$, $Q(\theta)=Q(\hat{\theta})=Q_k$ if $|\{i: \theta^i=\theta_h\}|=|\{i: \hat{\theta}^i=\theta_h\}|=k$. Moreover, it was shown that $Q_k=0$ if $k<k^*$. Here, it is further shown that if $(Q,P) \in \mathcal{M}$ is properly robust and $K \in \{k^*, \ldots, I\}$, then
$Q_K= q_K / q_I$, where $q_K$ is defined in the statement of Proposition \ref{thm_pub_proper}. By Proposition \ref{proper_leximin}, if a mechanism is properly robust, then it is leximin optimal. Because $Q_k=0$ for $k<k^*$, the lowest order statistics of the payoff vector are fixed at zero. Therefore, leximin optimality first requires maximizing the minimum payoff among $k \geq k^*$. That is, $Q_{k^*}, \ldots, Q_I$ must be chosen to maximize
\[\min_{k \in \{k^*, \ldots, I\}}Q_k S_k -k (\theta_h-\theta_\ell) Q_{k-1}\]
subject to $Q_{k^*-1}=0$ and the monotonicity constraint $0 \leq Q_{k^*} \leq \cdots \leq Q_I \leq 1$. This problem is equivalently formulated as
\begin{equation}
\begin{aligned}
&\max_{t,Q_{k^*}, \ldots, Q_I} t\\
&\subjecto\\
&Q_k S_k- k (\theta_h-\theta_\ell) Q_{k-1} \geq t \quad \text{for all $k=k^*, \ldots, I$,}\\
&Q_{k^*-1}=0, \quad \text{and}\\
&0 \leq Q_{k^*} \leq \cdots \leq Q_I \leq 1.
\end{aligned} \label{pub_lexi}
\end{equation}
For each $t \geq 0$, define the component-wise minimal sequence $(Q_{k^*}(t),\ldots,Q_I(t))$ satisfying the payoff inequalities in \eqref{pub_lexi} by
\[
Q_{k^*}(t)=\frac{t}{S_{k^*}}
\]
and, for $k>k^*$,
\[
Q_k(t)S_k-k(\theta_h-\theta_\ell)Q_{k-1}(t)=t,
\quad \text{or equivalently} \quad
Q_k(t)=\frac{t+k(\theta_h-\theta_\ell)Q_{k-1}(t)}{S_k}.
\]
Iterating this recursion yields
\[
Q_K(t)=tq_K, \quad \text{where} \quad 
q_K=\sum_{k=k^*}^K \left( \frac{1}{S_k}\prod_{j=k+1}^K \frac{j(\theta_h-\theta_\ell)}{S_j}\right)
\qquad\text{for } K\in\{k^*,\ldots,I\}.
\] 
Now let $(Q_{k^*},\ldots,Q_I)$ be any feasible solution to \eqref{pub_lexi} with objective value $t$. By construction of the minimal sequence, $Q_K\geq Q_K(t)$ for each $K\in\{k^*,\ldots,I\}$. In particular,
\[
Q_I\geq Q_I(t)=tq_I.
\]
Since feasibility also requires $Q_I\leq 1$, it follows that
\[
t\leq \frac{1}{q_I}.
\]
Moreover, equality holds if and only if $Q_I=1$ and $Q_K=Q_K(t)=q_K/q_I$ for every $K\in\{k^*,\ldots,I\}$. Thus, the conditions stated in the proposition are necessary.

\subsubsection{Sufficiency}\label{pub_sufficiency}

To prove that the mechanism defined in the proposition is properly robust, a full support and strongly adversarial LPS $\mu$ is constructed against which the mechanism is optimal. Define $\mu=(\mu_1, \mu_2)$ such that the support of $\mu_1$ is $\{\theta \in \Theta: \sum_i \theta^i \leq c\}$ and the support of $\mu_2$ is $\{\theta \in \Theta: \sum_i \theta^i>c\}$. For $k \geq k^*$, define $\Theta_k:= \{\theta \in \Theta: |\{i: \theta^i=\theta_h\}|=k\}$ and, for each $\theta \in \Theta_k$, set
\[ \mu_2(\theta)= \frac{1}{|\Theta_k|} \frac{w_k}{q_I}, \quad \text{where} \quad w_k:=\frac{1}{S_k} \prod^I_{j=k+1} \frac{j(\theta_h-\theta_\ell)}{S_j}.\] Note that $w_k S_k= (k+1) (\theta_h-\theta_\ell) w_{k+1}$. It is immediate that $\mu$ has full support. It is strongly adversarial because the mechanism defined in the proposition yields $0$ on $\{\theta \in \Theta: \sum_i \theta^i \leq c\}$ and $1/q_I>0$ on $\{\theta \in \Theta: \sum_i \theta^i>c\}$. So, type profiles in $\{\theta \in \Theta: \sum_i \theta^i \leq c\}$ are infinitely more likely than those in $\{\theta \in \Theta: \sum_i \theta^i>c\}$ and payoffs are lower in the former set.

It remains to verify $\mu$-optimality. Any $(Q,P) \in \mathcal{M}$ that maximizes expected utility with respect to $\mu_1$ must set $Q(\theta)=0$ if $\sum_i \theta^i<c$.
Because $\mu_2$ assigns equal weight to all profiles with the same number of $\theta_h$ types and the feasible set $\mathcal{M}$ is convex and permutation-invariant, for any feasible mechanism there exists an anonymous mechanism yielding the same $\mu_2$-expected payoff. Moreover, with transfers given by \eqref{rev_pub}, setting provision to zero at zero-surplus profiles preserves feasibility, leaves the first-order payoff unchanged, and weakly increases the second-order
payoff by the same argument used in the proof of Proposition \ref{thm_pub_perfect}. Hence, it suffices to restrict attention to anonymous allocation rules with transfers given by \eqref{rev_pub} and $Q_k=0$ for all $k< k^*$. Any such allocation rule is characterized by the increasing probability sequence $(Q_{k^*}, \ldots, Q_I)$ and yields a $\mu_2$-expected payoff of
\begin{equation}
    \begin{aligned}
\sum_{\theta \in \Theta} \mu_2(\theta) v((Q,P),\theta) &= \sum^I_{k=k^*} \sum_{\theta \in \Theta_k} \frac{1}{|\Theta_k|} \frac{w_k}{q_I} \left(Q_k S_k - k (\theta_h-\theta_\ell) Q_{k-1} \right) \\ &= \frac{1}{q_I} \sum^I_{k=k^*} w_k \left(Q_k S_k - k (\theta_h-\theta_\ell) Q_{k-1} \right)\\
&= \frac{1}{q_I} \left(\sum^I_{k=k^*} w_k S_k Q_k -\sum^{I-1}_{k=k^*}  \underbrace{(k+1) (\theta_h-\theta_\ell) w_{k+1}}_{=w_k S_k} Q_{k} \right)\\
&= \frac{1}{q_I} w_I S_I Q_I= \frac{Q_I}{q_I} \leq \frac{1}{q_I}.
    \end{aligned} \notag
\end{equation}
Because the mechanism defined in the proposition yields $1/q_I$ for all $\theta$ in the support of $\mu_2$, it must therefore yield $1/q_I$ against $\mu_2$. Hence, it must be $\mu$-optimal.

\subsection{Proof of Proposition \ref{thm_pub_asymptotic}}\label{proof_thm_pub_asymptotic}

By Proposition \ref{thm_pub_proper}, $f^I(1)=1$ for any number of agents $I \geq 2$. So, $f^I(x)$ converges to $1$ for $x=1$. Moreover, by Proposition \ref{thm_pub_proper}, $f^I(x)=0$ for all $I$ and any $x \leq \rho$. So, $f^I(x)$ converges to $0$ for any $x \in [0,\rho]$. It thus suffices to show that $f^I(x)$ converges to $0$ for all $x \in (\rho,1)$. By Proposition \ref{thm_pub_proper}, for any $K \in \{k^*, \ldots, I\}$,
\[q^I_I=\sum^I_{k=k^*} \left(\frac{1}{S_k} \prod^I_{j=k+1} \frac{j (\theta_h-\theta_\ell)}{S_j} \right) \geq \sum^K_{k=k^*} \left(\frac{1}{S_k} \prod^I_{j=k+1} \frac{j (\theta_h-\theta_\ell)}{S_j} \right)= \left(\prod^I_{j=K+1} \frac{j (\theta_h-\theta_\ell)}{S_j} \right) q^I_K. \] So,
\[ Q^I_K=\frac{q^I_K}{q^I_I} \leq \prod^I_{j=K+1} \frac{S_j}{j (\theta_h-\theta_\ell)}= \prod^I_{j=K+1} \left(1-\frac{\rho I}{j}\right) \leq \prod^I_{j=K+1} \left(1-\rho \right)= (1-\rho)^{I-K} .\] 
Now, fix $x \in (\rho,1)$. Then,
\[ \lim_{I \rightarrow \infty} f^I(x)= \lim_{I \rightarrow \infty} Q^I_{\lfloor xI \rfloor} \leq \lim_{I \rightarrow \infty} (1-\rho)^{I-\lfloor xI \rfloor}= 0,\] thereby establishing pointwise convergence to $f^\infty$.
For the uniform convergence claim, fix $\varepsilon \in (0,1)$. Note that, for any $x\in[0,1-\varepsilon]$, $\lfloor xI\rfloor \le (1-\varepsilon)I$. Therefore,
\[
I-\lfloor xI\rfloor \ge \varepsilon I.
\]
For $x \in [0,1-\varepsilon]$, $f^\infty(x)=0$. So, for $x \in [0,1-\varepsilon]$, it follows that
\[
|f^I(x)-f^\infty(x)|
=
f^I(x)
=
Q^I_{\lfloor xI\rfloor}
\le
(1-\rho)^{I-\lfloor xI\rfloor}
\le
(1-\rho)^{\varepsilon I}
=
\exp\!\big(-\varepsilon(-\ln(1-\rho))I\big).
\]
Defining $\lambda_\varepsilon:=\varepsilon(-\ln(1-\rho))$ and taking the supremum over $x\in[0,1-\varepsilon]$ yields
\[
\sup_{x\in[0,1-\varepsilon]} |f^I(x)-f^\infty(x)|
\le
\exp(-\lambda_\varepsilon I).
\]

\bibliography{references}

\addtocontents{toc}{\protect\setcounter{tocdepth}{0}}

\clearpage
\setcounter{page}{1}

\newcounter{daggerfootnote}
\newcommand*{\daggerfootnote}[1]{%
	\setcounter{daggerfootnote}{\value{footnote}}%
	\renewcommand*{\thefootnote}{\fnsymbol{footnote}}%
	\footnote[2]{#1}%
	\setcounter{footnote}{\value{daggerfootnote}}%
	\renewcommand*{\thefootnote}{\arabic{footnote}}%
}

\begin{center}
	\vspace*{2em}
	
	{\LARGE Supplemental Appendix for ``Lexicographic Robustness and the Efficiency
		of Optimal Mechanisms"\par}
	
	\vspace{1.5em}
	{\large Ashwin Kambhampati\daggerfootnote{
			Department of Economics, United States Naval Academy; \url{kambhamp@usna.edu}.}\par}
	
	\vspace{1em}
	{\large \today\par}
\end{center}

\vspace{1.5em}

\section{Supplemental Appendix}\label{app_examples}

\subsection{Screening: Example \ref{ex_robust}}\label{app_ex_screen_dom}

This appendix verifies that any robust mechanism that is not equal to the efficient and maximal mechanism is weakly dominated. Because any robust mechanism coincides with the efficient and maximal mechanism at $\theta_1$, the weak dominance claim is equivalent to the claim that, for any robust mechanism $(Q,P) \neq (Q^*,P^*)$, $v((Q^*,P^*),\theta_2) > v((Q,P),\theta_2)$.  Fixing $(q^*_1,p^*_1)$, it is straightforward to show that $(q^*_2, p^*_2)$ uniquely maximizes the seller's payoff from $\theta_2$ subject to the downward-adjacent incentive compatibility constraint, thereby satisfying the inequality. To see why, observe that stochastic mechanisms are strictly suboptimal in the example given the linearity of $u$ and the strict convexity of $c$. So, it suffices to show that $(q^*_2, p^*_2)$ is the unique solution to the relaxed problem
\begin{equation}
\begin{aligned}
&\max_{q_2 , p_2} \quad p_2-\frac{1}{2} q^2_2 \\
&\text{subject to}\\
& u(q_2, \theta_{2})-p_2 \geq u(q^*_1, \theta_{2})- p^*_1.
\end{aligned} \notag
\end{equation}
In any solution, the incentive compatibility constraint binds, yielding $p_2= u(q_2, \theta_2)+p^*_1-u(q^*_1, \theta_2)$. Eliminating constants from the objective function, it follows that any optimal quality must solve
\begin{equation}
\max_{q_2} \quad \theta_2 q_2 -\frac{1}{2} q^2_2. \notag
\end{equation}
The unique solution is $q_2=\theta_2$, which yields $p_2= \theta^2_1+\theta^2_2-\theta_2 \theta_1$. So, $(q^*_2, p^*_2)$ uniquely maximizes the seller's payoff from $\theta_2$.

\subsection{Screening: Example \ref{ex_properrobust}}\label{app_ex_properrobust}
This appendix verifies that $(Q,P) \in \mathcal{M}$ is $\mu$-optimal with respect to the adversarial and full support LPS $\mu=(\delta_1, \delta_3, \delta_2)$. First, observe that $(Q(\theta_1), P(\theta_1))$ uniquely attains the highest possible payoff against $\delta_1$. Second, from standard constraint simplification arguments, any mechanism that is full-information optimal for the lowest type cannot attain a payoff against $\delta_3$ higher than the value of the following relaxed problem:
\begin{equation}
\begin{aligned}
&\max_{(q_2,p_2), (q_3, p_3)} \quad p_3-\frac{1}{2} q^2_3 \\
&\text{subject to}\\
& u(q_3, \theta_{3})-p_3 \geq u(q_2, \theta_{3})-p_2\\
& u(q_2, \theta_{2})-p_2 \geq u(q^*_1, \theta_{2})- u(q^*_1, \theta_1)\\
& q_3 \geq q_2 \geq q^*_1.
\end{aligned} \notag
\end{equation}
It is immediate that, in any solution, the second incentive constraint binds. If not, then $p_2$ and $p_3$ can both be increased by the same
sufficiently small amount, preserving all constraints and strictly
increasing profit. So, $p_2= u(q^*_1, \theta_1)+ u(q_2, \theta_2)-u(q^*_1, \theta_2)$
in any solution. The relaxed problem thus reduces to
\begin{equation}
\begin{aligned}
&\max_{q_2, (q_3,p_3) } \quad p_3-\frac{1}{2} q^2_3 \\
&\text{subject to}\\
& u(q_3, \theta_{3})-p_3 \geq u(q_2, \theta_{3})- u(q_2, \theta_2)-\left( u(q^*_1, \theta_1) -u(q^*_1, \theta_2) \right)\\
& q_3 \geq q_2 \geq q^*_1.
\end{aligned} \notag
\end{equation}
Notice that the incentive constraint must bind in any solution. Moreover, the right-hand side of the incentive constraint is minimized subject to $q_2 \geq q^*_1$ if and only if $q_2=q^*_1$ because
$u(q_2, \theta_3)-u(q_2,\theta_2)= (\theta_3-\theta_2) q_2$. Hence, in any solution, it must be that $q_2=q^*_1$ and $p_3= u(q^*_1, \theta_1)+u(q_3, \theta_3)- u(q^*_1, \theta_3)$.
Substituting $p_3$ into the objective function and eliminating constants yields the surplus maximization problem
\begin{equation}
\max_{q_3} \quad \theta_3 q_3-\frac{1}{2} q^2_3, \notag
\end{equation}
whose solution is $q_3 = \theta_3$. It follows that $(Q,P)$ maximizes the seller’s second-order payoff (it is optimal in the relaxed problem and is feasible in the constrained problem). Among mechanisms maximizing the first- and second-order payoffs, the allocation following report $\theta_2$ has mean $\theta_1$ and price $\theta_1^2$. Strict convexity of the cost function implies that any nondegenerate allocation lottery following this report strictly reduces the seller’s third-order payoff. Therefore, $(Q,P)$ is uniquely optimal against $(\delta_1,\delta_3,\delta_2)$.

\subsection{Screening: Example \ref{ex_properrobust} continued}\label{app_ex_properrobust_2}

This appendix verifies that the efficient and maximal mechanism is $\mu$-optimal with respect to the full support and strongly adversarial LPS $\mu=(\delta_1, \delta_2, \delta_3)$. Let $(q^*_i, p^*_i)$ denote the quality-price pair of the good purchased by type $\theta_i$ in the efficient and maximal mechanism. Observe again that $(q^*_1, p^*_1)$ is uniquely optimal against $\delta_1$. Moreover, observe that any mechanism that is optimal against $\delta_1$ cannot attain a payoff against $\delta_2$ higher than the value of the following relaxed problem:
\begin{equation}
\begin{aligned}
&\max_{q_2 , p_2} \quad p_2-\frac{1}{2} q^2_2 \\
&\text{subject to}\\
& u(q_2, \theta_{2})-p_2 \geq u(q^*_1, \theta_{2})- p^*_1.
\end{aligned} \notag
\end{equation}
In any solution, the incentive constraint binds, yielding $p_2= u(q_2, \theta_2)+p^*_1-u(q^*_1, \theta_2)$. Eliminating constants from the objective function, it follows that any optimal quality must solve
\begin{equation}
\max_{q_2} \quad \theta_2 q_2 -\frac{1}{2} q^2_2. \notag
\end{equation}
The unique solution is $q_2=\theta_2$, which yields $p_2= u(q^*_1, \theta_1)+(u(q^*_2,\theta_2)-u(q^*_1, \theta_2))$. So, $(q^*_2, p^*_2)$ uniquely maximizes the seller's payoff. Finally, observe that any mechanism that is optimal against $\delta_2$ subject to optimality against $\delta_1$ cannot attain a payoff against $\delta_3$ higher than the value of the following relaxed problem:
\begin{equation}
\begin{aligned}
&\max_{q_3 , p_3} \quad p_3-\frac{1}{2} q^2_3 \\
&\text{subject to}\\
& u(q_3, \theta_{3})-p_3 \geq u(q^*_2, \theta_{3})- p^*_2.
\end{aligned} \notag
\end{equation}
In any solution, the incentive constraint binds, yielding $p_3= u(q_3, \theta_3)-u(q^*_2,\theta_3)+p^*_2$. Eliminating constants from the objective function, it follows that any optimal quality must solve
\begin{equation}
\max_{q_3} \quad \theta_3 q_3 -\frac{1}{2} q^2_3. \notag
\end{equation}
The unique solution is $q_3=\theta_3$, which yields $p_3= u(q^*_1, \theta_1)+(u(q^*_2,\theta_2)-u(q^*_1, \theta_2))+(u(q^*_3,\theta_3)-u(q^*_2,\theta_3))$. So, $(q^*_3, p^*_3)$ uniquely maximizes the seller's payoff. It follows that the efficient and maximal mechanism is $\mu$-optimal.

\subsection{Auctions: inefficient, weakly dominated robust mechanism }\label{app_ex_robust_auction}

This appendix identifies a robust auction mechanism that is inefficient and weakly dominated. Suppose there are two agents ($I=2$) and two types ($N=2$). Proposition \ref{thm_auction_robust} establishes that a mechanism $(Q,P) \in \mathcal{M}$ is robustly optimal if and only if
\[ Q^1(\underline{\theta})+Q^2(\underline{\theta}) =1, \quad
P^1(\underline{\theta})+P^2(\underline{\theta}) =\theta_1,\]
and the profitability constraints are satisfied:
\[P^1(\theta_1,\theta_2)+P^2(\theta_1,\theta_2) \geq \theta_1, \quad 
P^1(\theta_2,\theta_1)+P^2(\theta_2,\theta_1) \geq \theta_1, \quad \text{and} \quad 
P^1(\theta_2,\theta_2)+P^2(\theta_2,\theta_2) \geq \theta_1.\]
Let $(Q,P) \in \mathcal{M}$ be the mechanism in which $P$ is determined by \eqref{rev_auction} and $Q$ is given by
\begin{equation}
\begin{aligned}
\left(Q^1(\underline{\theta}),Q^2(\underline{\theta}) \right) &= \left(\frac{1}{2}, \frac{1}{2} \right), \\
\left(Q^1(\theta_1, \theta_2),Q^2(\theta_1, \theta_2) \right)&= \left(0, \frac{1}{2}+\frac{1}{2}\frac{\theta_1}{\theta_2} \right),\\
\left(Q^1(\theta_2, \theta_1),Q^2(\theta_2, \theta_1) \right)&= \left(\frac{1}{2}+\frac{1}{2} \frac{\theta_1}{\theta_2}, 0 \right), \quad \text{and}\\
\left(Q^1(\theta_2,\theta_2),Q^2(\theta_2,\theta_2)\right)&=\left(\frac{1}{2} \frac{\theta_1}{\theta_2}, \frac{1}{2} \frac{\theta_1}{\theta_2} \right).
\end{aligned} \notag
\end{equation}
By Proposition \ref{thm_auction_robust}, this mechanism is robustly optimal; it is efficient for the lowest type profile $\underline{\theta}$, $P^1(\underline{\theta})+P^2(\underline{\theta})=\theta_1$, and revenue is equal to $\theta_1$ at every type profile. Nevertheless, the mechanism is inefficient for all type profiles other than $\underline{\theta}$ because it does not allocate the good with probability one. Moreover, it is weakly dominated by the efficient and maximal mechanism with uniform tie-breaking.

\subsection{Auctions: withholding inefficiency}\label{app_ex_auction_ineff2}

This appendix identifies a perfectly robust auction mechanism that exhibits withholding inefficiency. Suppose there are two agents ($I=2$), three types ($N=3$), and $\theta_n=n$. Consider the mechanism $(Q,P) \in \mathcal{M}$ in which $P$ satisfies \eqref{rev_auction} and, for $i=1,2$ and $j \neq i$,
\begin{equation}
Q^i(\theta^i,\theta^j)=\begin{cases}
1 & \text{if $\theta^i>\theta^j$}\\
\frac{1}{2} & \text{if $\theta^i=\theta^j= \theta_1$ or $\theta^i=\theta^j= \theta_3$,}\\
\frac{1}{4} & \text{if $\theta^i=\theta^j = \theta_2$, and}\\
0 & \text{if $\theta^i<\theta^j$.}
\end{cases}
\notag
\end{equation} Note that $(Q,P)$ is inefficient because it allocates the good with probability $1/2$ under $(\theta_2,\theta_2)$. It is also perfectly robust; it is $\mu$-optimal with respect to the adversarial and full support LPS $\mu=(\delta_{\underline{\theta}}, \mu_2, \mu_3, \mu_4, \delta_{\overline{\theta}})$, where
\begin{equation}
\begin{aligned}
\mu_2 &= \frac{1}{5} \circ (\theta_2, \theta_2)+\frac{2}{5} \circ (\theta_3, \theta_2)+\frac{2}{5} \circ (\theta_2, \theta_3), \\
\mu_3 &= \frac{1}{2} \circ (\theta_1,\theta_2)+\frac{1}{2} \circ (\theta_2,\theta_1), \quad \text{and} \\
\mu_4 &= \frac{1}{2} \circ (\theta_1,\theta_3)+\frac{1}{2} \circ (\theta_3,\theta_1).
\end{aligned} \notag
\end{equation}  
The inefficiency arises because, under $\mu$, extracting rent from $(\theta_3,\theta_2)$ and $(\theta_2,\theta_3)$ is just as valuable to the seller as obtaining additional revenue at $(\theta_2,\theta_2)$. Note that $Q^1(\theta_2,\theta_2)+Q^2(\theta_2,\theta_2) \geq 1/2$ is necessary for $\mu$ to be adversarial with respect to the mechanism, though it will not be necessary to establish $\mu$-optimality.

To establish $\mu$-optimality, impose throughout the necessary condition for optimality that $P$ satisfies \eqref{rev_auction}. It is then immediate that a mechanism $(Q,P)\in\mathcal M$ maximizes the first-order payoff if and only if
\[
Q^1(\theta_1,\theta_1)+Q^2(\theta_1,\theta_1)=1.
\]
Now, fix any mechanism $(Q,P)\in\mathcal M$ that is optimal against $\delta_{\underline{\theta}}$.
To maximize the second-order payoff, it must solve
\begin{equation}
\begin{aligned}
& \max_{Q: \Theta \rightarrow [0,1]^2} \quad
\frac{2}{5} v((Q,P),(\theta_3,\theta_2))+\frac{2}{5} v((Q,P),(\theta_2,\theta_3))+\frac{1}{5} v((Q,P),(\theta_2,\theta_2)) \\
&\text{subject to}\\
&Q^1(\theta_1,\theta_1)+Q^2(\theta_1,\theta_1)=1\\
& Q^i(\cdot,\theta^{-i})~\text{increasing for all $i$ and $\theta^{-i} \in \Theta^{-i}$}\\
& \sum^2_{i=1} Q^i(\theta) \leq 1~\text{for all $i$ and $\theta \in \Theta$,}
\end{aligned}
\notag
\end{equation}
where
\begin{equation}
\begin{aligned}
v((Q,P),(\theta_3,\theta_2))=& Q^1(\theta_1,\theta_2)+\left(Q^1(\theta_2,\theta_2)-Q^1(\theta_1,\theta_2)\right)2+\left(Q^1(\theta_3,\theta_2)-Q^1(\theta_2,\theta_2)\right)3\\
& \qquad +Q^2(\theta_3,\theta_1)+\left(Q^2(\theta_3,\theta_2)-Q^2(\theta_3,\theta_1)\right)2,\\
v((Q,P),(\theta_2,\theta_3))=&
Q^2(\theta_2,\theta_1)+\left(Q^2(\theta_2,\theta_2)-Q^2(\theta_2,\theta_1)\right)2+\left(Q^2(\theta_2,\theta_3)-Q^2(\theta_2,\theta_2)\right)3\\
& \qquad +Q^1(\theta_1,\theta_3)+\left(Q^1(\theta_2,\theta_3)-Q^1(\theta_1,\theta_3)\right)2, \quad \text{and}\\
v((Q,P),(\theta_2,\theta_2))=&Q^1(\theta_1,\theta_2)+\left(Q^1(\theta_2,\theta_2)-Q^1(\theta_1,\theta_2)\right)2\\
& \qquad +Q^2(\theta_2,\theta_1)+\left(Q^2(\theta_2,\theta_2)-Q^2(\theta_2,\theta_1)\right)2.
\end{aligned} \notag
\end{equation}
It is immediate that any solution must satisfy
\[
Q^1(\theta_1,\theta_2)=Q^1(\theta_1,\theta_3)=Q^2(\theta_3,\theta_1)=Q^2(\theta_2,\theta_1)=0 \quad \text{and} \quad Q^1(\theta_3,\theta_2)=Q^2(\theta_2,\theta_3)=1.
\]
Substituting these equalities into the objective function yields a constant. Therefore, no further restrictions are required for optimality. Now, fix any mechanism $(Q,P)\in\mathcal M$ that is $\mu$-optimal against the first two beliefs. To maximize the third-order
payoff, it must solve
\begin{equation}
\begin{aligned}
&\max_{Q: \Theta \rightarrow [0,1]^2}\quad
\frac{1}{2}\Bigl[v((Q,P),(\theta_1,\theta_2))\Bigr]
+\frac{1}{2}\Bigl[v((Q,P),(\theta_2,\theta_1))\Bigr]\\
&\text{subject to}\\
&Q^1(\theta_1,\theta_2)=Q^1(\theta_1,\theta_3)=Q^2(\theta_3,\theta_1)=Q^2(\theta_2,\theta_1)=0\\
& Q^1(\theta_3,\theta_2)=Q^2(\theta_2,\theta_3)=1\\
&Q^1(\theta_1,\theta_1)+Q^2(\theta_1,\theta_1)=1\\
& Q^i(\cdot,\theta^{-i})~\text{increasing for all $i$ and $\theta^{-i} \in \Theta^{-i}$}\\
& \sum^2_{i=1} Q^i(\theta) \leq 1~\text{for all $i$ and $\theta \in \Theta$,}
\end{aligned}
\notag
\end{equation}
where
\[v((Q,P),(\theta_1,\theta_2))=Q^1(\theta_1,\theta_2)+Q^2(\theta_1,\theta_1)+\left(Q^2(\theta_1,\theta_2)-Q^2(\theta_1,\theta_1)\right)2\]
and
\[v((Q,P),(\theta_2,\theta_1))=Q^1(\theta_1,\theta_1)+\left(Q^1(\theta_2,\theta_1)-Q^1(\theta_1,\theta_1)\right)2+Q^2(\theta_2,\theta_1).\]
Imposing $Q^1(\theta_1,\theta_2)=Q^2(\theta_2,\theta_1)=0$ and $Q^1(\theta_1,\theta_1)+Q^2(\theta_1,\theta_1)=1$, the objective function reduces to
\begin{equation}
-\frac{1}{2}
+Q^2(\theta_1,\theta_2)+Q^1(\theta_2,\theta_1).
\notag
\end{equation}
Hence, any solution must satisfy
\[
Q^2(\theta_1,\theta_2)=Q^1(\theta_2,\theta_1)=1.
\]
Notice that monotonicity immediately implies
\[
Q^2(\theta_1,\theta_3)=Q^1(\theta_3,\theta_1)=1.
\]
Hence, any mechanism that is $\mu$-optimal with respect to the first three beliefs is also $\mu$-optimal with respect to the first four beliefs. The fifth-order problem remains:
\begin{equation}
\begin{aligned}
&\max_{Q: \Theta \rightarrow [0,1]^2}\quad 3\bigl(Q^1(\theta_3,\theta_3)+Q^2(\theta_3,\theta_3)\bigr)\\
&\text{subject to}\\
&Q^1(\theta_1,\theta_2)=Q^1(\theta_1,\theta_3)=Q^2(\theta_3,\theta_1)=Q^2(\theta_2,\theta_1)=0\\
& Q^2(\theta_1,\theta_3)=Q^1(\theta_3,\theta_1)=Q^2(\theta_1,\theta_2)=Q^1(\theta_2,\theta_1)=Q^1(\theta_3,\theta_2)=Q^2(\theta_2,\theta_3)=1\\
&Q^1(\theta_1,\theta_1)+Q^2(\theta_1,\theta_1)=1\\
& Q^i(\cdot,\theta^{-i})~\text{increasing for all $i$ and $\theta^{-i} \in \Theta^{-i}$}\\
& \sum^2_{i=1} Q^i(\theta) \leq 1~\text{for all $i$ and $\theta \in \Theta$.}
\end{aligned}
\notag
\end{equation}
Any allocation rule satisfying the constraints that sets
\[
Q^1(\theta_3,\theta_3)+Q^2(\theta_3,\theta_3)=1
\]
is optimal. In summary, it has been shown that any mechanism $(Q,P) \in \mathcal{M}$ in which $P$ satisfies \eqref{rev_auction} and $Q$ satisfies
\begin{equation}
\begin{aligned}
&Q^1(\theta_1,\theta_2)=Q^1(\theta_1,\theta_3)=Q^2(\theta_3,\theta_1)=Q^2(\theta_2,\theta_1)=0,\\
& Q^2(\theta_1,\theta_3)=Q^1(\theta_3,\theta_1)=Q^2(\theta_1,\theta_2)=Q^1(\theta_2,\theta_1)=Q^1(\theta_3,\theta_2)=Q^2(\theta_2,\theta_3)=1,\\
&Q^1(\theta_3,\theta_3)+Q^2(\theta_3,\theta_3)=Q^1(\theta_1,\theta_1)+Q^2(\theta_1,\theta_1)=1,\\
& Q^i(\cdot,\theta^{-i})~\text{increasing for all $i$ and $\theta^{-i} \in \Theta^{-i}$}, \quad \text{and}\\
& \sum^2_{i=1} Q^i(\theta) \leq 1~\text{for all $i$ and $\theta \in \Theta$}
\end{aligned}
\notag
\end{equation}
is $\mu$-optimal. In particular, the mechanism under consideration, which sets
$Q^1(\theta_2,\theta_2)=Q^2(\theta_2,\theta_2)=\frac{1}{4}$,
is $\mu$-optimal.

\subsection{Auctions: misallocation inefficiency}\label{app_ex_auction_misallocation}

This appendix identifies a perfectly robust auction mechanism that exhibits misallocation inefficiency. Suppose there are two agents ($I=2$) and three types ($N=3$).
Consider the mechanism $(Q,P)\in\mathcal M$ in which $P$ satisfies \eqref{rev_auction}
and the allocation rule $Q$ is defined as follows:
\[
Q^1(\theta^1,\theta^2)=
\begin{cases}
1 & \text{if $\theta^1>\theta^2$ and $(\theta^1,\theta^2)\neq (\theta_2,\theta_1)$}\\
0 & \text{otherwise}
\end{cases}
\qquad \text{and} \qquad
Q^2(\theta^1,\theta^2)=1-Q^1(\theta^1,\theta^2).
\]
Notice that ties are always broken in favor of agent $2$. Moreover, under the type profile
$(\theta_2,\theta_1)$, the good is allocated to agent $2$ even though agent $1$ has a strictly higher value. Hence, $(Q,P)$ is inefficient: the good is sometimes allocated to a bidder whose value is strictly smaller than that of the other bidder. It is also perfectly robust; it is
$\mu$-optimal with respect to the full support and adversarial LPS
\[
\mu=(\delta_{\underline{\theta}},\ \delta_{(\theta_3,\theta_1)},\ \delta_{(\theta_3,\theta_2)},
\ \mu_4,\ \mu_5),
\]
where
\[
\mu_4 = \frac{1}{3} \circ (\theta_1,\theta_2)+\frac{1}{3} \circ (\theta_1,\theta_3)+\frac{1}{3} \circ (\theta_2,\theta_1)
\]
and
\[\mu_5 = \frac{1}{3} \circ (\theta_2,\theta_2)+\frac{1}{3} \circ (\theta_2, \theta_3)+\frac{1}{3} \circ (\theta_3,\theta_3).
\] The inefficiency arises because, to maximize revenue at $(\theta_3,\theta_1)$, it is optimal to distort agent $1$'s allocation downward completely whenever agent $2$'s type is $\theta_1$ and agent $1$'s type is strictly less than $\theta_3$. This leads to inefficiency at $(\theta_2, \theta_1)$.

To establish $\mu$-optimality, impose the necessary condition for optimality that
$P$ satisfies \eqref{rev_auction}. It is then immediate that a mechanism $(Q,P)\in\mathcal M$
maximizes the first-order payoff if and only if
\[
Q^1(\theta_1,\theta_1)+Q^2(\theta_1,\theta_1)=1.
\]
Now, fix any mechanism $(Q,P)\in\mathcal M$ that is optimal against $\delta_{\underline{\theta}}$.
To maximize the second-order payoff, it must solve
\begin{equation}
\begin{aligned}
& \max_{Q:\Theta\to [0,1]^2} \quad v((Q,P),(\theta_3,\theta_1))\\
&\text{subject to}\\
&Q^1(\theta_1,\theta_1)+Q^2(\theta_1,\theta_1)=1\\
& Q^i(\cdot,\theta^{-i})~\text{increasing for all $i$ and $\theta^{-i}\in\Theta^{-i}$}\\
& \sum_{i=1}^2 Q^i(\theta)\leq 1~\text{for all $\theta\in\Theta$,}
\end{aligned}
\notag
\end{equation}
where
\begin{equation}
\begin{aligned}
v((Q,P),(\theta_3,\theta_1))
=&\; Q^1(\theta_1,\theta_1) \theta_1
+\bigl(Q^1(\theta_2,\theta_1)-Q^1(\theta_1,\theta_1)\bigr)\theta_2
+\bigl(Q^1(\theta_3,\theta_1)-Q^1(\theta_2,\theta_1)\bigr)\theta_3\\
&\qquad + Q^2(\theta_3,\theta_1)\theta_1.
\end{aligned}
\notag
\end{equation}
It is immediate that any solution must satisfy
\[
Q^1(\theta_3,\theta_1)=1
\qquad \text{and} \qquad
Q^2(\theta_3,\theta_1)=Q^1(\theta_2,\theta_1)=Q^1(\theta_1,\theta_1)=0.
\]
By the first-order optimality condition, it then follows that
\[
Q^2(\theta_1,\theta_1)=1.
\]
Monotonicity then implies
\[Q^2(\theta_1,\theta_2)=Q^2(\theta_1,\theta_3)=1.\]
Hence, by feasibility,
\[Q^1(\theta_1,\theta_3)=Q^1(\theta_1,\theta_2)=0.\]
Now, fix any mechanism $(Q,P)\in\mathcal M$ that is $\mu$-optimal against the first two beliefs.
To maximize the third-order payoff, it must solve
\begin{equation}
\begin{aligned}
& \max_{Q:\Theta\to [0,1]^2} \quad v((Q,P),(\theta_3,\theta_2))\\
&\text{subject to}\\
&Q^1(\theta_1,\theta_3)=Q^1(\theta_1,\theta_2)=Q^2(\theta_3,\theta_1)=Q^1(\theta_2,\theta_1)=Q^1(\theta_1,\theta_1)=0\\
&Q^2(\theta_1,\theta_2)=Q^2(\theta_1,\theta_3)=Q^1(\theta_3,\theta_1)=Q^2(\theta_1,\theta_1)=1\\
& Q^i(\cdot,\theta^{-i})~\text{increasing for all $i$ and $\theta^{-i}\in\Theta^{-i}$}\\
& \sum_{i=1}^2 Q^i(\theta)\leq 1~\text{for all $\theta\in\Theta$,}
\end{aligned}
\notag
\end{equation}
where
\begin{equation}
\begin{aligned}
v((Q,P),(\theta_3,\theta_2))
=&\; Q^1(\theta_1,\theta_2)\theta_1
+\bigl(Q^1(\theta_2,\theta_2)-Q^1(\theta_1,\theta_2)\bigr)\theta_2
+\bigl(Q^1(\theta_3,\theta_2)-Q^1(\theta_2,\theta_2)\bigr)\theta_3\\
&\qquad + Q^2(\theta_3,\theta_1)\theta_1
+\bigl(Q^2(\theta_3,\theta_2)-Q^2(\theta_3,\theta_1)\bigr)\theta_2.
\end{aligned}
\notag
\end{equation}
Imposing $Q^1(\theta_1,\theta_2)=Q^2(\theta_3,\theta_1)=0$ and collecting terms yields
\[
v((Q,P),(\theta_3,\theta_2))
=
\theta_3 Q^1(\theta_3,\theta_2)+\theta_2 Q^2(\theta_3,\theta_2)
-(\theta_3-\theta_2)Q^1(\theta_2,\theta_2).
\]
It is immediate that any solution must satisfy
\[
Q^1(\theta_3,\theta_2)=1
\quad \text{and} \quad
Q^2(\theta_3,\theta_2)=Q^1(\theta_2,\theta_2)=0.
\]
Now, fix any mechanism $(Q,P)\in\mathcal M$ that is $\mu$-optimal against the first three beliefs. To maximize the fourth-order payoff, $Q$ must solve
\begin{equation}
\begin{aligned}
& \max_{Q:\Theta\to [0,1]^2} \quad \frac{1}{3} v((Q,P),(\theta_1,\theta_2))+\frac{1}{3} v((Q,P), (\theta_1, \theta_3)) + \frac{1}{3} v((Q,P), (\theta_2, \theta_1))\\
&\text{subject to}\\
&Q^2(\theta_3,\theta_2)=Q^1(\theta_2,\theta_2)=Q^1(\theta_1,\theta_3)=Q^1(\theta_1,\theta_2)=Q^2(\theta_3,\theta_1)=Q^1(\theta_2,\theta_1)=Q^1(\theta_1,\theta_1)=0\\
&Q^1(\theta_3,\theta_2)=Q^2(\theta_1,\theta_2)=Q^2(\theta_1,\theta_3)=Q^1(\theta_3,\theta_1)=Q^2(\theta_1,\theta_1)=1\\
& Q^i(\cdot,\theta^{-i})~\text{increasing for all $i$ and $\theta^{-i}\in\Theta^{-i}$}\\
& \sum_{i=1}^2 Q^i(\theta)\leq 1~\text{for all $\theta\in\Theta$,}
\end{aligned}
\notag
\end{equation}
where
\begin{equation}
\begin{aligned}
v((Q,P),(\theta_1,\theta_2))
=&\; Q^1(\theta_1,\theta_2)\theta_1+Q^2(\theta_1,\theta_1)\theta_1
+\bigl(Q^2(\theta_1,\theta_2)-Q^2(\theta_1,\theta_1)\bigr)\theta_2,\\
v((Q,P),(\theta_1,\theta_3))
=&\; Q^2(\theta_1,\theta_1)\theta_1
+\bigl(Q^2(\theta_1,\theta_2)-Q^2(\theta_1,\theta_1)\bigr)\theta_2
+\bigl(Q^2(\theta_1,\theta_3)-Q^2(\theta_1,\theta_2)\bigr)\theta_3 
\\& \quad +Q^1(\theta_1,\theta_3)\theta_1, \quad \text{and}\\
v((Q,P),(\theta_2,\theta_1))
=&\; Q^1(\theta_1,\theta_1)\theta_1
+\bigl(Q^1(\theta_2,\theta_1)-Q^1(\theta_1,\theta_1)\bigr)\theta_2+Q^2(\theta_2,\theta_1)\theta_1.
\end{aligned}
\notag
\end{equation}
Imposing $Q^1(\theta_2,\theta_1)=Q^1(\theta_1,\theta_1)=Q^1(\theta_1,\theta_3)=Q^1(\theta_1,\theta_2)=0$ and $Q^2(\theta_1,\theta_3)=Q^2(\theta_1,\theta_2)=Q^2(\theta_1,\theta_1)=1$ the objective function reduces to
\[\frac{1}{3} \theta_1+ \frac{1}{3} \theta_1+\frac{1}{3} Q^2(\theta_2,\theta_1) \theta_1.\]
It is immediate that the allocation rule maximizes the fourth-order payoff if and only if it satisfies the constraints and
\[Q^2(\theta_2,\theta_1)=1.\]
Monotonicity then implies
\[Q^2(\theta_2,\theta_2)=Q^2(\theta_2,\theta_3)=1.\]
Hence, by feasibility,
\[Q^1(\theta_2,\theta_3)=0.\]
Finally, fix any mechanism $(Q,P)\in\mathcal M$ that is $\mu$-optimal against the first four beliefs. The only remaining free variables are $Q^1(\theta_3,\theta_3)$ and $Q^2(\theta_3,\theta_3)$. These values enter into the auctioneer's objective function under $\mu_5$ only through $v((Q,P),(\theta_3,\theta_3))$. Hence, to maximize the fifth-order payoff, it suffices to solve
\begin{equation}
\begin{aligned}
& \max_{Q^1(\theta_3,\theta_3), Q^2(\theta_3,\theta_3) \geq 0} \quad \frac{1}{3}(Q^1(\theta_3,\theta_3)+Q^2(\theta_3,\theta_3))\theta_3\\
&\text{subject to}\\
& \sum_{i=1}^2 Q^i(\theta)\leq 1~\text{for all $\theta\in\Theta$.}
\end{aligned}
\notag
\end{equation}
Any $Q^1(\theta_3,\theta_3), Q^2(\theta_3,\theta_3) \geq 0$ such that $Q^1(\theta_3,\theta_3)+Q^2(\theta_3,\theta_3)=1$ solves this problem. Observe that the allocation rule considered satisfies this condition. Hence, it is $\mu$-optimal.

\subsection{Public good: perfect robustness example}\label{app_ex_pub}
This appendix illustrates the lack of discipline imposed by perfect robustness in the public good model. Suppose there are two agents ($I=2$) and $c \in (2 \theta_\ell, \theta_h+ \theta_\ell)$. Proposition \ref{thm_pub_robust} and Proposition \ref{thm_pub_perfect} together imply that, in any perfectly robust mechanism $(Q,P) \in \mathcal{M}$, $P$ satisfies \eqref{rev_pub}, $Q(\underline{\theta})=0$, $Q(\overline{\theta})=1$, and
\[Q(\theta_h,\theta_\ell)+Q(\theta_\ell,\theta_h) \leq \frac{2 \theta_h-c}{\theta_h-\theta_\ell},\] where the final condition ensures that the firm's ex post payoff is nonnegative under type profile $\overline{\theta}$. It turns out that these necessary conditions for perfect robustness are, in fact, sufficient for perfect robustness in this example. To see why, fix any such mechanism $(Q,P) \in \mathcal{M}$ and consider the adversarial and full support LPS \[\mu=(\delta_{\underline{\theta}},\alpha \circ (\theta_h,\theta_\ell)+\alpha \circ (\theta_\ell, \theta_h)+(1-2\alpha) \circ \overline{\theta}),\] where
\[\alpha:= \left(2+\frac{\theta_h+\theta_\ell-c}{\theta_h-\theta_\ell} \right)^{-1}.\] Impose the necessary condition for optimality that
$P$ satisfies \eqref{rev_pub}. It is then immediate that a mechanism $(Q,P)\in\mathcal M$
maximizes the first-order payoff if and only if
\[
Q(\underline{\theta})=0.
\] Now, fix any mechanism $(Q,P)\in\mathcal M$ that is optimal against $\delta_{\underline{\theta}}$.
To maximize the second-order payoff, it must solve
\begin{equation}
\begin{aligned}
&\max_{Q:\Theta\to [0,1]} \quad \alpha v((Q,P),(\theta_h,\theta_\ell))
+\alpha v((Q,P),(\theta_\ell,\theta_h))
+(1-2\alpha)v((Q,P),\overline{\theta})\\
&\text{subject to}\\
&Q(\underline{\theta})=0\\
&Q(\cdot,\theta^2)~\text{and}~Q(\theta^1,\cdot)~\text{increasing},
\end{aligned}
\notag
\end{equation}
where
\begin{equation}
\begin{aligned}
v((Q,P),(\theta_h,\theta_\ell))
&=(\theta_h+\theta_\ell-c)Q(\theta_h,\theta_\ell),\\
v((Q,P),(\theta_\ell,\theta_h))
&=(\theta_h+\theta_\ell-c)Q(\theta_\ell,\theta_h), \quad \text{and}\\
v((Q,P),\overline{\theta})
&=(2\theta_h-c)Q(\overline{\theta})
-(\theta_h-\theta_\ell)\bigl(Q(\theta_h,\theta_\ell)+Q(\theta_\ell,\theta_h)\bigr).
\end{aligned}
\notag
\end{equation}
Using the definition of $\alpha$, the coefficients on
$Q(\theta_h,\theta_\ell)$ and $Q(\theta_\ell,\theta_h)$ cancel, so the problem becomes
equivalent to
\[
\max_{Q:\Theta\to [0,1]} \quad (1-2\alpha)(2\theta_h-c)Q(\overline{\theta}).
\]
Since $1-2\alpha>0$ and $2\theta_h-c>0$, any solution must satisfy
\[
Q(\overline{\theta})=1.
\]
It has thus been shown that any mechanism $(Q,P)\in\mathcal M$ in which $P$
satisfies \eqref{rev_pub} and $Q$ satisfies
\[
Q(\underline{\theta})=0
\quad \text{and} \quad
Q(\overline{\theta})=1
\]
is $\mu$-optimal. Because the mechanism in the example satisfies these conditions, it is therefore $\mu$-optimal.

\subsection{Calculations for Figure \ref{fig_converge}}\label{app_fig_converge}

This appendix presents the calculations used for Figure \ref{fig_converge}. Figure \ref{fig_converge} sets $\theta_h=1$, $\theta_\ell=0$, and
$\gamma=1/2$, so that $S_k=k-I/2$ and
$k^*=\lfloor I/2\rfloor+1$.
For each $I\in\{5,15,55\}$, the formula in
Proposition \ref{thm_pub_proper} is evaluated using the recursion
\[
q_{k^*-1}=0 \quad \text{and} \quad
q_k=\frac{1+kq_{k-1}}{k-I/2}
\quad\text{for }k=k^*,\ldots,I.
\]
The provision probability is then $Q^I_k=0$ for $k<k^*$ and
$Q^I_k=q_k/q_I$ otherwise. Table \ref{tab_figure_coordinates} reports the non-zero provision
probabilities, rounded to five significant digits. The corresponding
horizontal coordinate is $x=k/I$.

\begin{table}[htbp]
	\centering
	\caption{Positive provision probabilities in Figure \ref{fig_converge}.}
	\label{tab_figure_coordinates}
	\small
	\setlength{\tabcolsep}{4pt}
	\renewcommand{\arraystretch}{1.1}
	\begin{tabular}{rr@{\hspace{1.2em}}rr@{\hspace{1.2em}}rr}
		\hline
		\multicolumn{2}{c}{$I=5$}
		& \multicolumn{2}{c}{$I=15$}
		& \multicolumn{2}{c}{$I=55$} \\
		$k$ & $Q^5_k$ & $k$ & $Q^{15}_k$
		& $k$ & $Q^{55}_k$ \\
		\hline
		3 & $0.16129$ & 8 & $0.00045778$ & 28 & $1.5266\times10^{-15}$ \\
		4 & $0.48387$ & 9 & $0.0028993$ & 29 & $3.0022\times10^{-14}$ \\
		5 & $1$ & 10 & $0.011689$ & 30 & $3.6057\times10^{-13}$ \\
		& & 11 & $0.036801$ & 31 & $3.1939\times10^{-12}$ \\
		& & 12 & $0.098187$ & 32 & $2.2712\times10^{-11}$ \\
		& & 13 & $0.23212$ & 33 & $1.3627\times10^{-10}$ \\
		& & 14 & $0.49998$ & 34 & $7.1281\times10^{-10}$ \\
		& & 15 & $1$ & 35 & $3.3264\times10^{-9}$ \\
		& & & & 36 & $1.4088\times10^{-8}$ \\
		& & & & 37 & $5.4871\times10^{-8}$ \\
		& & & & 38 & $1.9858\times10^{-7}$ \\
		& & & & 39 & $6.7345\times10^{-7}$ \\
		& & & & 40 & $0.0000021550$ \\
		& & & & 41 & $0.0000065449$ \\
		& & & & 42 & $0.000018958$ \\
		& & & & 43 & $0.000052592$ \\
		& & & & 44 & $0.00014025$ \\
		& & & & 45 & $0.00036063$ \\
		& & & & 46 & $0.00089671$ \\
		& & & & 47 & $0.0021613$ \\
		& & & & 48 & $0.0050606$ \\
		& & & & 49 & $0.011533$ \\
		& & & & 50 & $0.025630$ \\
		& & & & 51 & $0.055622$ \\
		& & & & 52 & $0.11806$ \\
		& & & & 53 & $0.24537$ \\
		& & & & 54 & $0.50000$ \\
		& & & & 55 & $1$ \\
		\hline
	\end{tabular}
\end{table}

\end{document}